\documentclass[aps,prl,twocolumn,superscriptaddress]{revtex4-1}
\usepackage{graphicx}  
\usepackage{amsmath} 
\usepackage{amssymb}
\usepackage{enumerate}
\pdfoutput=1

\newcommand{\Eq}[1]{Eqn.~\ref{#1}}

\newcommand{\Fig}[1]{Fig.~\ref{#1}}

\newcommand{\eq}{\begin{equation}}
\newcommand{\qe}{\end{equation}}
\newcommand{\pvec}[1]{\vec{#1}\mkern2mu\vphantom{#1}}

\usepackage{color}
\begin{document}

\title{Tunable Band Topology in Gyroscopic Lattices}
\author{Noah P. Mitchell}
\email{npmitchell@uchicago.edu}
\thanks{Corresponding author}
\affiliation{James Franck Institute and Department of Physics, University of Chicago, Chicago, IL 60637, USA}
\author{Lisa M. Nash}
\affiliation{James Franck Institute and Department of Physics, University of Chicago, Chicago, IL 60637, USA}
\author{William T. M. Irvine}
\email{wtmirvine@uchicago.edu}
\thanks{Corresponding author}
\affiliation{James Franck Institute and Department of Physics, University of Chicago, Chicago, IL 60637, USA}
\affiliation{Enrico Fermi Institute, The University of Chicago, Chicago, IL 60637, USA}

\begin{abstract}
Gyroscopic metamaterials --- mechanical structures composed of interacting spinning tops --- have recently been found to support one-way topological edge excitations.
In these structures, the time reversal symmetry breaking that enables their topological behavior emerges directly from the lattice geometry.
Here we show that variations in the lattice geometry can therefore give rise to more complex band topology than has been previously described.
A `spindle' lattice (or truncated hexagonal tiling) of gyroscopes possesses both clockwise and counterclockwise edge modes distributed across several band gaps. 
Tuning the interaction strength or twisting the lattice structure along a Guest mode opens and closes these gaps and yields bands with Chern numbers of $|C| > 1$ without introducing next-nearest-neighbor interactions or staggered potentials.
A deformable honeycomb structure provides a simple model for understanding the role of lattice geometry in constraining the effects of time reversal symmetry and inversion symmetry breaking.
Lastly, we find that topological band structure generically arises in gyroscopic networks, and a simple protocol generates lattices with topological excitations. 

\end{abstract}

\maketitle

\section{Introduction}
Materials with nontrivial band topology have captured the attention of condensed matter scientists since their discovery in electronic systems~\cite{thouless_quantized_1982}.
Since then, the concept of topological order has found its way to a plethora of physical systems, from electronic to
photonic, acoustic, and even mechanical systems~\cite{prodan_topological_2009,rechtsman_photonic_2013,nash_topological_2015,mitchell_realization_2018,mitchell_amorphous_2018,wang_topological_2015,kane_topological_2013,susstrunk_observation_2015,haldane_possible_2008,ningyuan_time-_2015,peano_topological_2015,wang_reflection-free_2008,fleury_floquet_2016}.
When topologically nontrivial, all these systems exhibit excitations confined to their surface that propagate unidirectionally without backscattering and are robust to disorder. 
These features are both fundamentally intriguing and form the basis for technological applications of topological materials.

In mechanical systems, these chiral edge waves have recently been demonstrated using structures composed of coupled gyroscopes~\cite{nash_topological_2015,wang_topological_2015,mitchell_realization_2018,mitchell_amorphous_2018}. 
To date, studies of topological gyroscopic materials have focused on a handful of simple lattices that possess a nonzero Chern numbers in their phononic band structure.
The minimal requirements for such a Chern insulator are the the presence of a band gap and broken time reversal symmetry.
In the electronic case, time reversal symmetry breaking arises from the presence of magnetic fields~\cite{haldane_model_1988}. 
As we will see, the analogous mechanism in gyroscopic lattices is the lattice geometry itself: the mere presence of spinning components is not sufficient to generate the effects enabling chiral edge modes.

In this article, we go beyond simple geometries and find the flexibility to design lattices with desired band gaps and desired topology. 
In particular, we examine tunable lattices with Chern numbers $|C| > 1$ as well as multiple gaps with edge modes of opposite chirality, we examine the effects of competing time reversal symmetry breaking with inversion symmetry breaking, and demonstrate a design strategy to achieve band topology in lattices with seemingly arbitrary unit cells.

\begin{figure}[ht]

\includegraphics[]{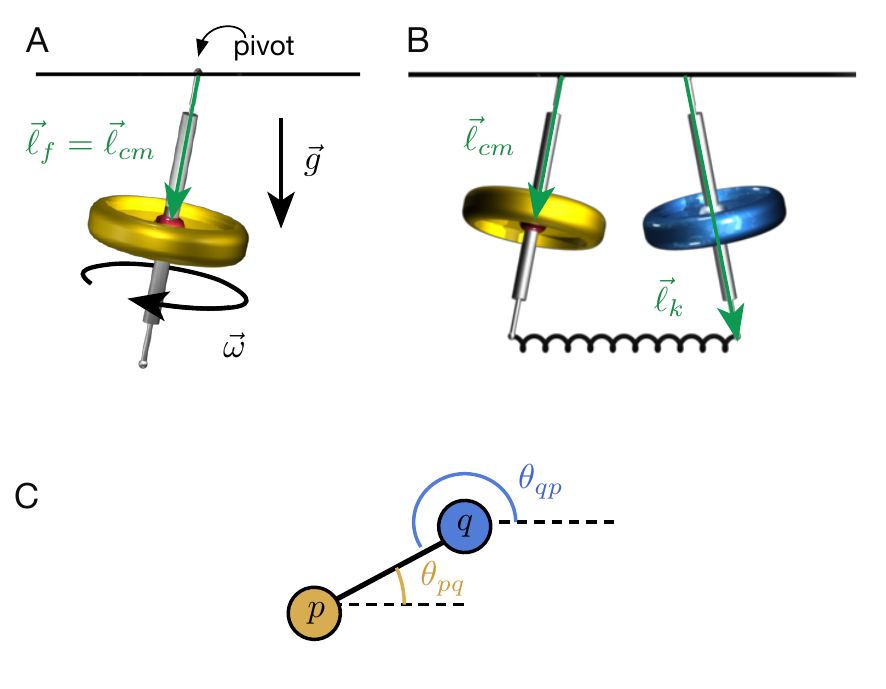}
\caption[]{
\textbf{
A spring-coupled gyroscopic metamaterial is composed of spinning gyroscopes that hang from a pinned pivot point.}
 \textit{(A)} 
 $\vec{\ell_f}$ is the vector from the pivot point to where a force acts.
When the only force is gravity, $\vec{\ell_f} = \vec{\ell_{cm}}$  
 \textit{(B)} Gyroscopes in the metamaterial are coupled to their neighbors in the lattice via a spring which is attached to the free end.  
 \textit{(C)} The linearized equation of motion for our system relates the displacements via angles between bonds and the local gyroscope's local $x$-axis (indicated by dotted lines in this view from above).
 }
\label{gyro_figure}
\end{figure}


\section{The equations of motion}

A simple realization of gyroscopic metamaterials is a collection of coupled gyroscopes which hang from a pivot point and spin rapidly enough for their angular momentum to lie approximately along the primary axis, as shown in \Fig{gyro_figure}A. 
Under these conditions, the free tip of a gyroscope moves when a torque, $\vec{\tau}$, acts about the pivot point according to: 
\begin{equation}
	\label{eq:simpEOM}
	 \vec{\tau} \approx I\omega \dot{\hat{n}} 
     = \vec{\ell_f} \times \vec{F} 	
\end{equation}
where $\vec{\ell_f}$ as the vector from the pivot point to the point acted upon by force, $\vec{F}$.

Considering small displacements of each gyroscope allows a linearized description. 
Denoting the displacement from the equilibrium position in the plane as $\psi = x + i y$,  the equation of motion for a single gyroscope under the influence of gravity becomes: 
\begin{equation}
	\label{grav}
	\dot{\psi} = i \frac{m g \ell_{cm}}{I \omega}\psi.
\end{equation}
Note that throughout this paper, without loss of generality, we choose the angular momentum vector of a hanging gyroscope to point down.
Noting the similarity between Equation~\ref{grav} and the Schr\"{o}dinger equation for a quantum particle, we use the same notion of time reversal symmetry as is used in quantum mechanics, namely $\psi \rightarrow \psi^*$ and $t \rightarrow -t$. 
While $\psi \rightarrow \psi^*$ corresponds to a reversal of momentum for a quantum particle, in the context of gyroscopes, $\psi \rightarrow \psi^*$ carries out a reflection of the gyroscope's displacement about a horizontal axis passing through its pivot point.
Performing this operation on the equation above, 
we find that Equation~\ref{grav} is time reversal symmetric. 
Thus, a spinning top precessing under the influence of gravity does not break this notion of time reversal symmetry.

Introducing interactions, however, allows the structure to break time reversal symmetry.
The simplest setting to see this is a network of gyroscopes coupled by linear springs.
For small displacements, the forces exerted on one gyroscope by another are proportional to the component of the net displacement along the line connecting them.
For a given pair of gyroscopes $p$ and $q$, it is convenient to extract the component of the net displacement $\psi_p - \psi_q$ along the bond by rotating the system to the local $x-$axis of $p$, taking the real part of expression, and then rotating back. 
The resulting force in complex form is given by:
\begin{align}
F_{pq} &= -k_0 e^{i \theta_{pq}} \textrm{Re}[e^{-i \theta_{pq}}(\psi_p - \psi_q)] \\
&=-\frac{k_0 e^{i \theta_{pq}}}{2} \left[e^{-i \theta_{pq}}(\psi_p - \psi_q) + e^{i \theta_{pq}}(\psi^*_p-\psi^*_q)\right], \nonumber
\end{align}  
where $k_0$ is the spring constant of the bond.
Using this result, the equation of motion for two gyroscopes can then be written as:
\begin{equation}
	\label{twoeom}
	\begin{split}
	i \dot{\psi}_p =
    -\bigg( \frac{ \Omega_k}{2}  \bigg[ &\left(\psi_p - \psi_q\right)
	\\&+  e^{2i \theta_{pq}}\left(\psi^*_p - \psi^*_q\right)\bigg] 
	 + \Omega_{g} \psi_p\bigg).
	\end{split}
\end{equation}
where $\Omega_k ={k_0 \ell_e^2}/({I \omega})$ and $\Omega_{g} = {mg\ell_{cm}/} (I \omega)$.
We define the time reversal operation as $\psi_p^{TR}(t) = \psi_p^* (-t)$.
By taking the complex conjugate of $\dot{\psi_p}$ and rewriting in terms of $\psi_p^{TR}$, we see that the equations of motion are changed only by $e^{i2\theta} \rightarrow e^{-i2\theta}$.
Therefore, we see that time reversal symmetry is preserved under reflections that are parallel or perpendicular to the bond~\cite{nash_topological_2015}.

The full equation of motion for a hanging gyroscope with more than one neighbor can be similarly expressed: 
\begin{equation}
	\label{eq_lattice_eom}
	\begin{split}
	i \dot{\psi}_p = -\bigg(\frac{ \Omega_k}{2} \sum_q^{n.n.}  \bigg[&\left(\psi_p - \psi_q\right) +  e^{2 i \theta_{pq}}\left(\psi^*_p - \psi^*_q\right)\bigg] 
	\\ & + \Omega_{g} \psi_p\bigg).
	\end{split}
\end{equation}
As before, time reversal symmetry is only preserved if all bonds are either parallel or perpendicular to each other since in this case, a coordinate system can be chosen so that bonds lie along the $x$ and $y$ axes, constraining the prefactor $e^{2i\theta_{pq}}$ to be real for all bonds in the network.

To date, the only ordered lattices that have been considered in this framework are the honeycomb lattice and simple distortions thereof~\cite{nash_topological_2015,mitchell_realization_2018}.
A slightly different manifestation of gyroscopic metamaterials considered in~\cite{wang_topological_2015} found that by including staggered sublattice precession frequencies and bond strengths, time reversal symmetry could be effectively broken in lattices with square and honeycomb symmetries.

\section{Twisted spindle lattice}

\begin{figure}[t!]
\begin{centering}
\includegraphics[width=90mm]{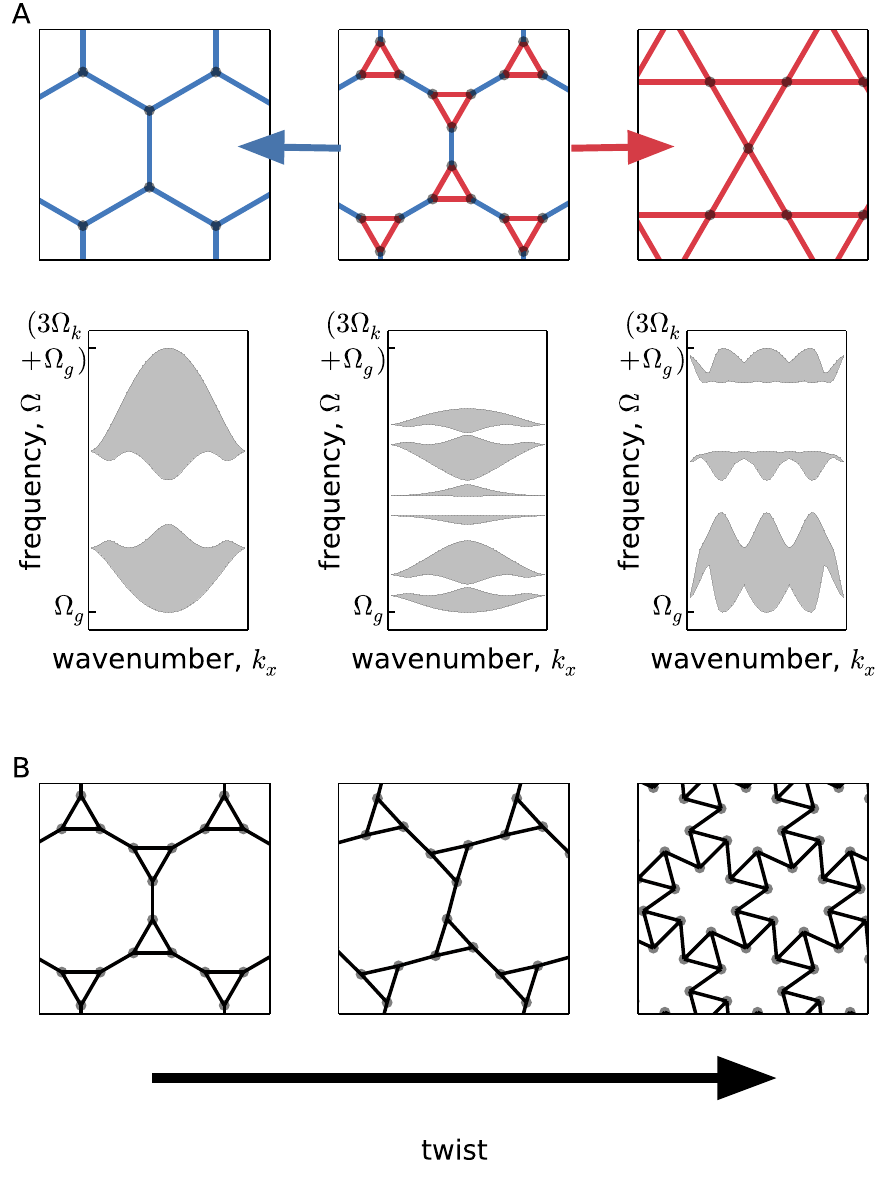}
\caption{
\textbf{The spindle lattice shares features of both the honeycomb and kagome lattices, while supporting a Guest mode in which each triad of gyroscopes is locally rotated.}
\textit{(A)} Taking the size of the red triangles in the spindle lattice to zero returns a honeycomb configuration, while taking the length of the blue bonds connecting each red triangle of gyroscopes to zero transforms the spindle lattice into the kagome configuration.
The associated band structures are shown below each of the three configurations.
\textit{(B)} Locally twisting the triangles spindle lattice preserves bond lengths while globally deforming the lattice.}
\label{spindle_geomintro}
\end{centering}
\end{figure}

To demonstrate the considerable flexibility of gyroscopic metamaterials, we begin by considering the twisted spindle lattice shown in \Fig{spindle_geomintro}.
This structure shares features of both the honeycomb lattice and the kagome lattice.
As \Fig{spindle_geomintro}A shows, shrinking the red triangles to a single site --- while increasing the strength of red bonds --- deforms the spindle lattice into the honeycomb lattice.
Conversely, taking the length of the blue bonds that connect triads of gyroscopes to zero --- while increasing their strength ---  deforms the spindle lattice into a kagome configuration. 
As shown in \Fig{spindle_geomintro}B, the spindle lattice also supports a Guest mode in which each triad of gyroscopes rotates locally.

\begin{figure*}[ht]
\includegraphics[width=\textwidth]{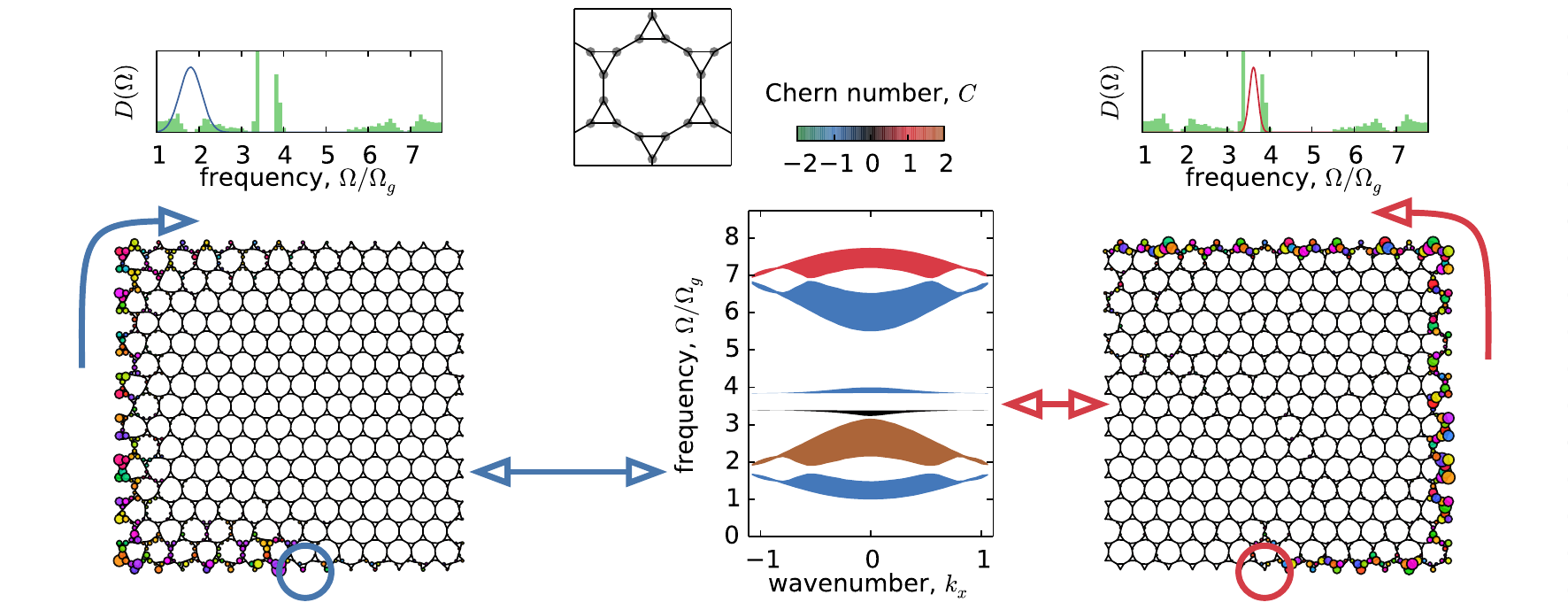}
\caption[]{
\textbf{The gyroscopic spindle lattice contains chiral edge modes of either chirality as well as bands with Chern number of $C > 1$.}
Direct simulation of Equation~\ref{eq_lattice_eom} reveal clockwise (left) and counterclockwise (right) edge modes in the same structure when shaken at different frequencies ($\omega = 1.8\, \Omega_g$ (left) and $3.62 \,\Omega_g$ (right)).
Computing the Chern numbers for each band confirms the topological origin of the chiral edge modes, as shown in the colored band structure in the middle panel. 
A single gyroscope on the edge is shaken at a fixed frequency with an amplitude varying in time; the spectrum of the excitation is indicated by blue (left) and red (right) curves overlaying the density of states, $D(\Omega)$.
The density of states, shown above each lattice, is given for the case with periodic boundary conditions.
For these simulations, the interaction strength was set to be $\Omega_k = 3 \Omega_g$.
}
\label{spindle_intro}
\end{figure*}
\begin{figure*}[ht]
\includegraphics[]{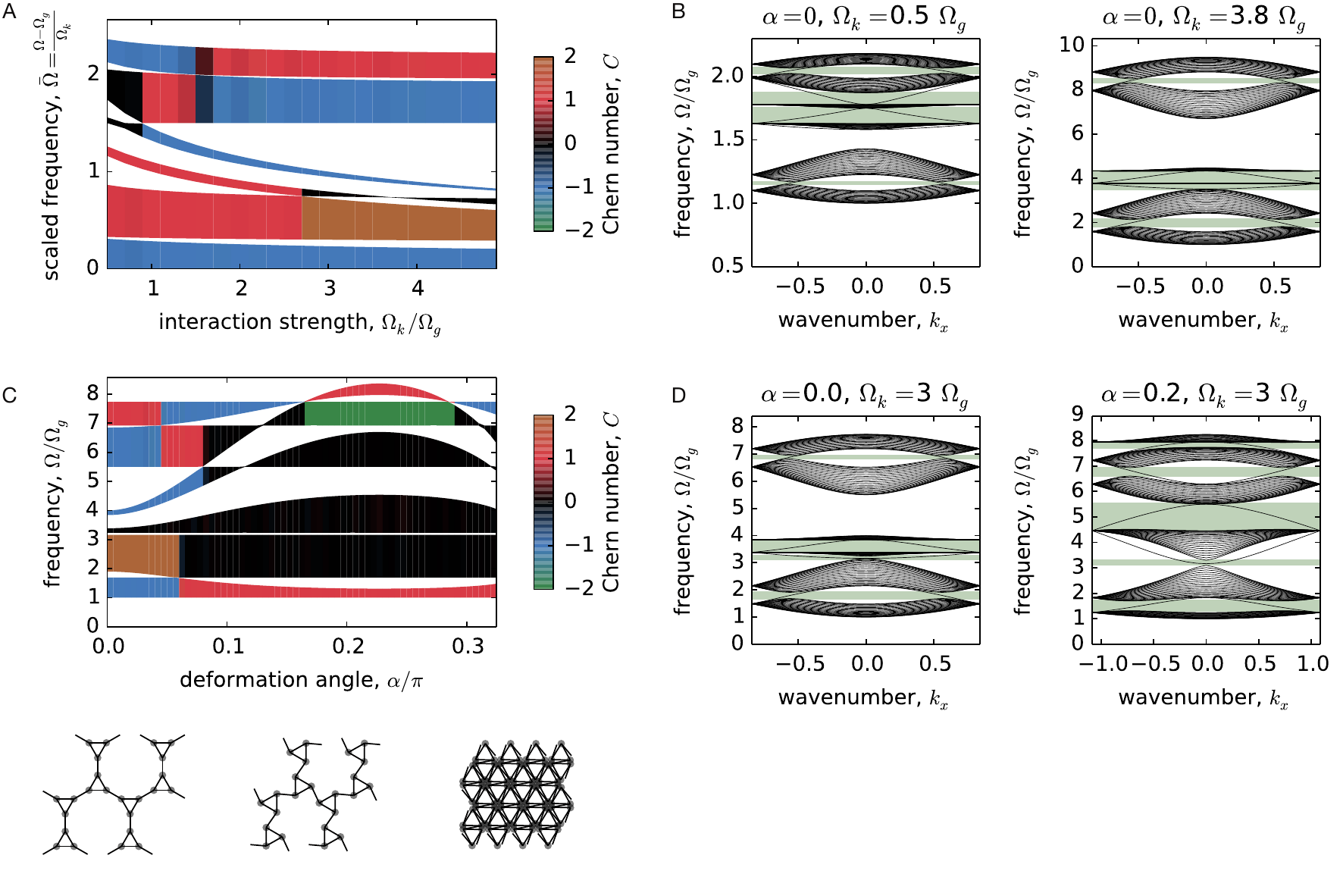}
\caption[]{
\textbf{Phononic band structures for the spindle and twisted spindle lattices show opening and closing of gaps with topological edge states.}
\textit{(A)} Simply increasing the interaction strength enables the closing and opening of band gaps, creating and annihilating protected chiral edge modes.
\textit{(B)} Band gaps for two different interaction strengths are highlighted in green.
\textit{(C-D)} As the structure is twisted through a bond-length-preserving Guest mode, three of the five gaps close and reopen, leading to three or four gaps with chiral edge modes, depending on the value of the twist deformation angle, $\alpha$. 
}
\label{spindle_bands}
\end{figure*}

\begin{figure*}[t]
\includegraphics[]{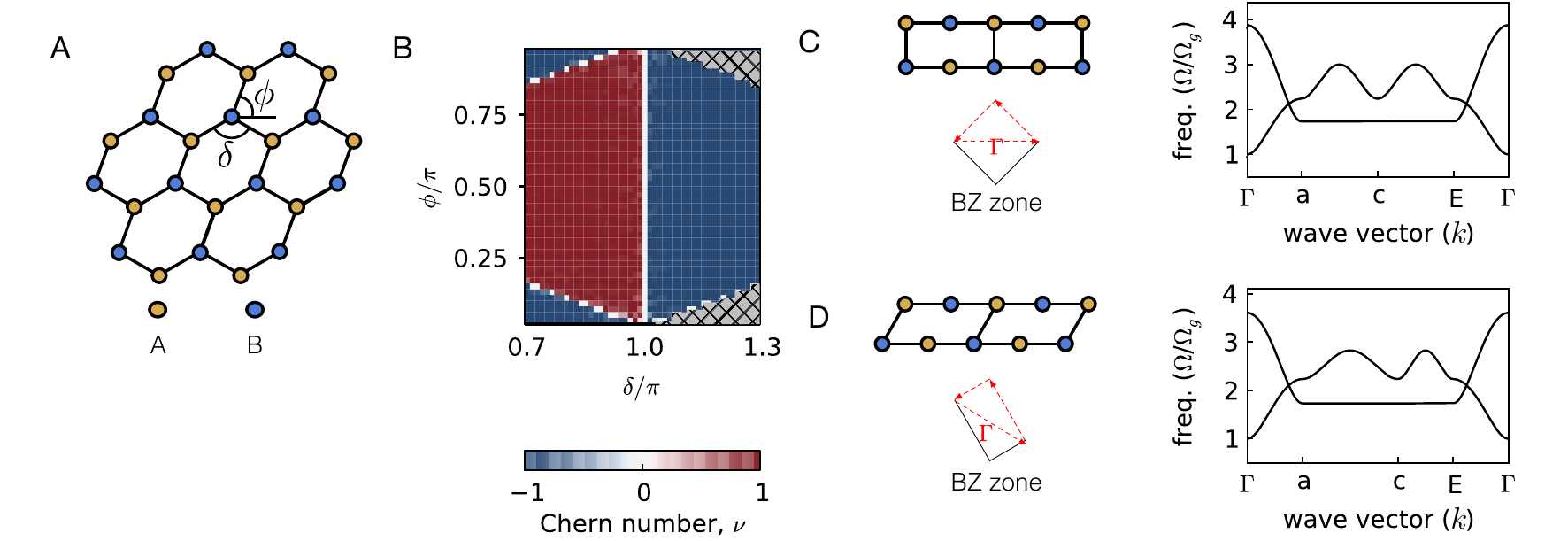}
\caption[]{\textbf{Band gaps and topology in the deformed honeycomb lattice} 
\textit{(A)} The angles $\phi$ and $\delta$ control the deformation of the honeycomb lattice. 
\textit{(B)} The $\phi$-$\delta$ phase diagram shows that the bulk Chern number the system changes when straight lines of bonds appear in the lattice, which occurs on the white diagonal lines in the left corners and on the white vertical line at $\delta = \pi$. 
\textit{(C)} The bricklayer configuration ($\delta = \pi$) band structure is plotted along paths in the Brillouin zone.  
The gap is closed at two Dirac points. 
\textit{(D)} No band gap opens in the canted bricklayer ($\delta = \pi$, $\phi \neq \pi/2 $), even though time reversal symmetry is broken in this configuration. 
}
\label{fig:honeycomb_bulk}
\end{figure*}

\begin{figure*}
\includegraphics[width=\textwidth]{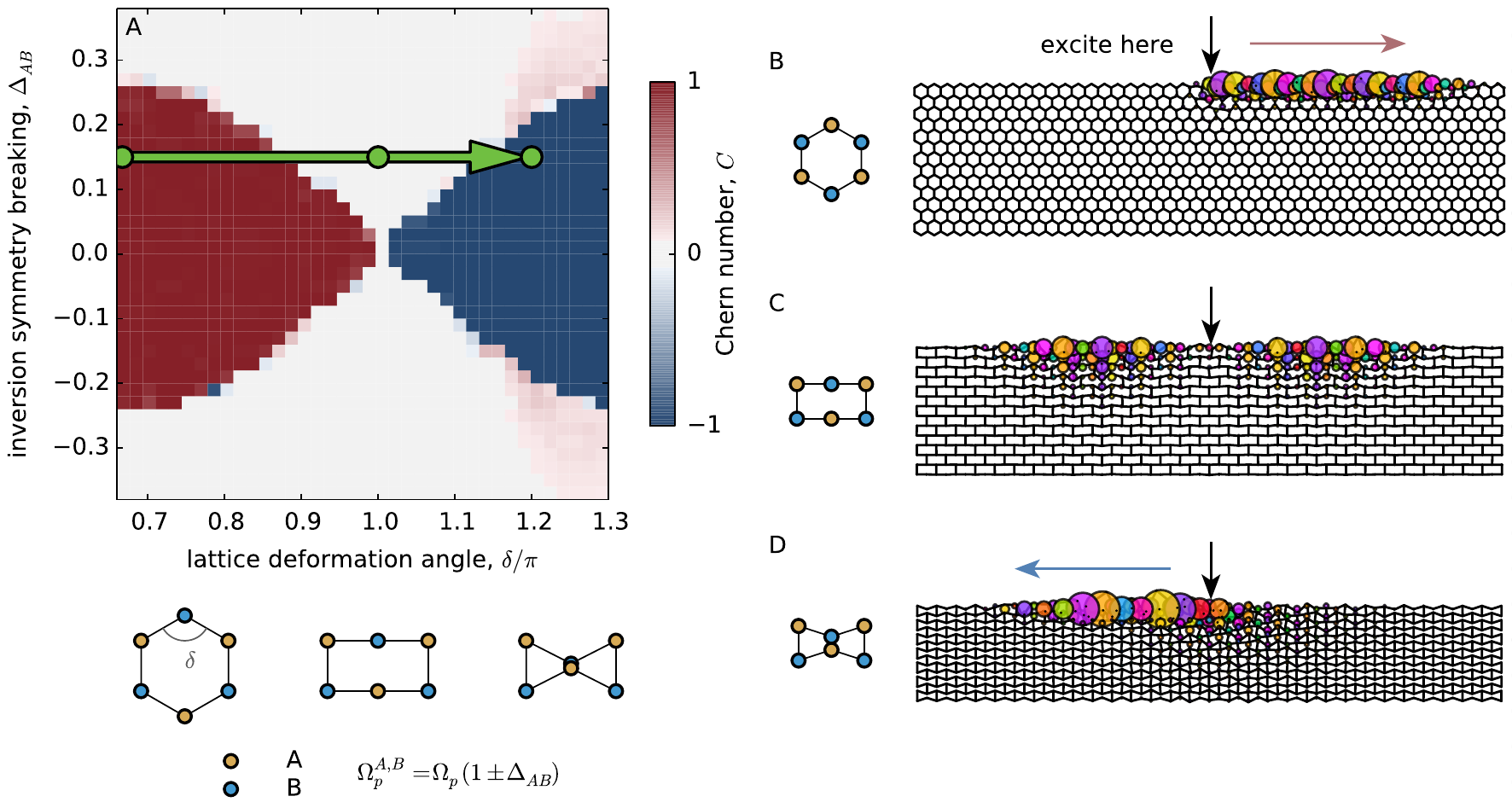}
\caption[]{
\textbf{Inversion and time reversal symmetries compete in a gyroscopic lattice.}
\textit{(A)} The phase diagram for a deformed honeycomb lattice (without shear, so that $\phi = \pi/2$) with varying $\Omega_g$ values on sites A and B shows an interplay between inversion and time reversal symmetries.
\textit{(B)} In a simulation of the honeycomb lattice with inversion symmetry breaking $\Delta_{AB} = 0.15$, driving a gyroscope on the edge at a gap frequency results in a clockwise wavepacket. 
\textit{(C)} When the lattice is deformed to a bricklayer geometry, the Chern number vanishes. 
This configuration is gapped due to the inversion symmetry breaking ($\Delta_{AB}=0.15$). 
The gap contains modes which are localized on the edge, but these unprotected edge waves propagate in both directions and are not robust against disorder. 
(See also Supplementary Videos 3-4.)
\textit{(D)} In the bowtie geometry, edge modes propagate counterclockwise, as predicted by the calculations shown in \textit{(A)}.}
\label{phase_diagram_delta}
\end{figure*}

In the limiting case of the honeycomb lattice, which has two sites per unit cell, we find a single gap with clockwise topologically-protected edge modes~\cite{nash_topological_2015}. 
By contrast, in the kagome lattice, with three sites per unit cell, there are two gaps, which each support a counterclockwise topological mode, as shown in the Supplementary Information~\cite{noauthor_notitle_nodate}.
In the intermediate case of the undeformed spindle lattice, which has six sites per unit cell, we generically find five band gaps. Most of these gaps possess chiral edge modes, and a given configuration can host both clockwise and counterclockwise modes. As we show below, locally twisting this structure (as in \Fig{spindle_geomintro}B) or varying the bond strengths, $\Omega_k$, relative to the pinning strength, $\Omega_g$, opens and closes edge-mode-carrying gaps.

Shaking a gyroscope on the boundary of this network at a frequency in the lowest band gap generates a clockwise wavepacket confined to the edge of the sample which is robust to disorder in the gravitational precession frequencies or bond strengths and does not scatter at sharp corners or defects (\Fig{spindle_intro}A and Supplementary Video 1~\cite{noauthor_notitle_nodate}). 
Shaking at a frequency in the middle band gap, however, generates counterclockwise edge waves, allowing a single lattice structure to conduct protected edge waves with a chirality determined by frequency (\Fig{spindle_intro}B).
We compute the Chern number for each band via~\cite{avron_homotopy_1983}
\begin{equation}
\label{chern_num}
C_{j} dx \wedge dy = \frac{i}{2 \pi} \int d^2 k\ \mathrm{Tr}[dP_j \wedge P_j dP_j],
\end{equation}
where $P_j \equiv |u_j \rangle \langle u_j |$ is the projection matrix and $\wedge$ is the wedge product, 
and we find that the Chern number is equal to the number of chiral edge modes, which suggests the same bulk-boundary correspondence for these systems as in electronic Chern insulators~\cite{laughlin_quantized_1981}.

The topological band structure of the gyroscopic spindle lattice offers additional axes of tunability through varying the interaction strength (ie, the bond stiffness in the case of springs) and by performing bond-length-preserving deformations on the lattice.
For gyroscopic lattices with uniform interaction strengths (ie. equal spring constants throughout), we can tune the 
the ratio of interaction frequency to gravitational precession frequency, $\Omega_k / \Omega_g$. 
This operation can deform the band structure in addition to changing its extent.
In the case of the spindle lattice, 
this provides a tuning knob that changes the topology of the band structure.
Simply increasing the interaction strength relative to the gravitational precession frequency closes and reopens gaps and changes the Chern numbers of bands, as shown in \Fig{spindle_bands}A.
We note that this feature was absent in the gyroscopic honeycomb lattice previously studied~\cite{nash_topological_2015,mitchell_realization_2018},
whose topology was unaffected by changes in $\Omega_k$ and $\Omega_g$. 
This allowed the topology to be continuously connected to the electronic Haldane model, unlike in the spindle lattice.

Twisting the spindle lattice through a Guest mode, as shown in \Fig{spindle_geomintro}B, also provides a tuning knob. 
Globally deforming the lattice closes and reopens the lowest and highest band gaps, allowing for several distinct configurations of multiple gaps supporting protected chiral edge modes, as shown in \Fig{spindle_bands}C and Supplementary Video 2~\cite{noauthor_notitle_nodate}.
As the twist angle grows, there are five values for which a pair of bands touch and reopen, flipping the chirality of the modes in that gap or imparting chiral modes to a gap which previously had none.

What determines the chirality of edge modes? 
Unlike in Maxwell lattices recently found to be topological~\cite{kane_topological_2013}, here the coordination number alone does not play a central role in determining band topology.
If Chern numbers were determined purely at the level of nearest neighbors, we would expect, for instance, that the spindle and honeycomb lattices would have similar edge modes:
both have a coordination number of $z=3$ and have obtuse bond angles $\theta_{nml} > \pi / 2$ for all junctions. 
However, the spindle supports edge modes of either chirality. 
Furthermore, the spindle lattice's rich band structure depends not only on geometry, but also bond strengths (\Fig{spindle_intro}A).
We conclude that simple, local aspects of the lattice such as coordination number and mean bond angle do not singlehandedly determine the band structure.

From a design perspective, the two simple tuning parameters of angle and interaction strength are sufficient to cover a broad range of topological phenomenology without introducing staggered interaction strengths, including edge modes with either chirality, the opening and closing of gaps, and bands with Chern number $\pm 1$ and $\pm 2$. These behaviors demonstrate the versatility of gyroscopic metamaterials.

\section{Time reversal symmetry and topological band-gaps}
All configurations shown so far break time reversal symmetry, which is a necessary ingredient for band topology in Chern insulators~\cite{nash_topological_2015,wang_topological_2015}.
This is not necessarily true for all gyroscopic lattices. 
For example, as illustrated in \Fig{fig:honeycomb_bulk}, a honeycomb lattice can undergo a bond-length-preserving deformation to a configuration in which all bond angles are multiples of $\pi/2$ (for $\delta=\pi$).
In such a configuration, time reversal symmetry is restored and therefore band topology disappears. 
Further changing the value of $\delta$ past $\pi$ causes the band topology to reappear, but with opposite sign.
In \Fig{fig:honeycomb_bulk}, we extend this analysis to the entire phase-space of periodic, bond-length-preserving deformations by introducing an additional angle, $\phi$.
This allows us to explore the question of whether time reversal symmetry breaking is sufficient to generate band topology in gyroscopic metamaterials.

\Fig{fig:honeycomb_bulk}B shows the topological phase diagram corresponding to general deformations of the honeycomb lattice, characterized by angles $\phi$ and $\delta$. 
Red (blue) regions indicate to a Chern number of 1 (-1) for the lower band and, correspondingly, clockwise (counter-clockwise) propagating modes in the gap.
For $\phi=\pi/2$ and $\delta=\pi$, the network is arranged in a bricklayer configuration (\Fig{fig:honeycomb_bulk}C).
Varying either $\phi$ or $\delta$ from this point breaks time reversal symmetry. 
However, only changes in $\delta$ imbue nontrivial band topology, as illustrated by the white line in \Fig{fig:honeycomb_bulk}B for $\delta=\pi$~\cite{nash_topological_2015}. 
The fact that changes in $\phi$ break time reversal symmetry without opening a gap demonstrates that broken time reversal symmetry does not inevitably lead to either band gaps or nontrivial band topology.

This behavior warrants further investigation.
During the deformation of the honeycomb into the bricklayer geometry, the band gap closes and the two Dirac points in the spectrum touch at a point. 
Surprisingly, these Dirac points are preserved even in the canted bricklayer configuration, as shown in \Fig{fig:honeycomb_bulk}D, despite the fact that shearing the bricklayer configuration breaks time reversal symmetry by creating acute and obtuse bond angles (see Equation~\ref{eq_lattice_eom}).

As detailed in the Supplementary Information, this protection of the Dirac cones arises due to a subtle pseudo-reflection symmetry. 
The symmetry consists of reflecting the positions of gyroscopes about the $x$ axis and shearing their relative positions such that the tilt angle $\phi$ is invariant, while leaving the gyroscopes' displacements unchanged.
This pseudo-reflection is a symmetry of the equations of motion, and thus of the normal modes~\cite{noauthor_notitle_nodate}. 
This symmetry leads to the existence of a special line of modes in momentum space. 
Along this line, modes that are symmetric and antisymmetric under the symmetry operation decouple and cannot hybridize at their band crossing.
Thus, a pseudo-reflection symmetry stabilizes the Dirac points against acquiring gaps, which would otherwise be unstable to time reversal symmetry breaking perturbations.
This symmetry also explains the vanishing Chern number for all values of $\phi$ at $\delta=\pi$ seen in \Fig{fig:honeycomb_bulk}B, on account of the Berry curvature being odd under the action of the symmetry.
More broadly, this protection underscores of the interplay between lattice geometry and the topological character of the band structure.

\section{Competing symmetries in topological gyroscopic systems}
Breaking inversion symmetry is the canonical mechanism for opening gaps in the phonon spectra of mass-and-spring lattices~\cite{chaikin_principles_2000}.
This is also true in other systems, such as electronic materials.
This gap opening mechanism can be made to compete with broken time reversal symmetry to close and reopen gaps and eliminate protected chiral edge modes.
To study an analogous effect in gyroscopic lattices, we detune the yellow and blue sublattice sites in \Fig{phase_diagram_delta} by modulating their on-site gravitational precession frequencies: $\Omega_{gA,B} = (1 \pm \Delta)\Omega_g$ (see also~\cite{mitchell_realization_2018}).

\Fig{phase_diagram_delta}A shows the phase diagram that results from varying $\delta$ and lattice pinning frequencies, $\Omega_{gA,B}$.
When the unit cell's two sites are equivalent ($\Delta = 0$), the Chern number of the system changes only when the gap closes at the bricklayer transition.
For $\Delta \neq 0$, however, a third, topologically trivial region appears.
In this case, the band structure is gapped, yet displays no chiral edge modes.

The behavior of excitations confirms the Chern number calculations in all three regions, as indicated in \Fig{phase_diagram_delta}B-D and Supplementary Video 3~\cite{noauthor_notitle_nodate}. 
In \Fig{phase_diagram_delta}C, excitations propagate along the edge in both directions.
The Chern number is zero, and these edge waves are not topologically protected: they backscatter at sharp corners or in the presence of disorder (see Supplementary Video 4~\cite{noauthor_notitle_nodate}).
The result shown in \Fig{phase_diagram_delta}A displays a strong resemblance to Haldane's phase diagram: sites must have similar pinning strengths for the lattice to support topological states.

\begin{figure}[tb]
\includegraphics[width=\columnwidth]{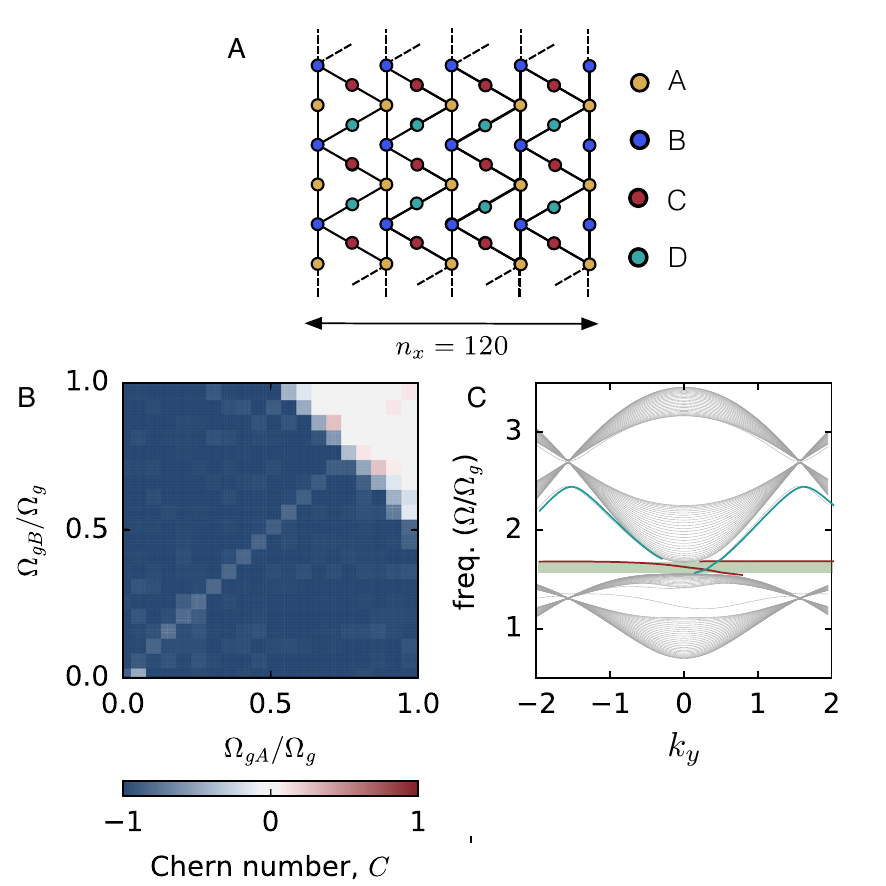}
\caption[]{\textbf{Coordination number and topological phases}.  \textit{(A)} A lattice with four lattice sites per unit cell, where sites A and B have two neighbors and sites C and D have two.
\textit{(B)} The topological phase diagram for varying the gravitational precession frequencies on sites A and B shows that because of the different coordination numbers for the lattice sites, the band structure is trivial when $\Omega_{gA} =\Omega_{gB}=\Omega_{gC}=\Omega_{gD}$.   
\textit{(C)} The band structure in the nontrivial phase for a strip which is infinite along $y$ and 120 unit cells wide in $x$.  
}
\label{fig:alpha_lattice}
\end{figure}

While varying precession frequencies is an effective way of breaking inversion symmetry, it is not the only one. An alternative way is to alter the coordination number between sites -- i.e. the number of bonds that are linked to each gyroscope.
For example, unlike the lattices considered so far, the $\alpha$--$(ET)_2 I_3$ lattice shown in \Fig{fig:alpha_lattice} contains sites of coordination number $z=4$ (for the $A$ and $B$ sites) and $z=2$ (for the $C$ and $D$ sites).
When all gravitational precession frequencies are equal, the lattice displays no topological excitations (top right corner of \Fig{fig:alpha_lattice}B).

As seen in the first term of \Eq{eq_lattice_eom}, contributions to on-site pinning --- ie. terms in which $\dot{\psi_p}$ depends on $\psi_p$ --- come not only from gravitation precession terms ($\Omega_g$), but also from coupling to adjacent sites.
For lattices with unequal coordination at different sites, balancing the full `site pinning frequency', $\Omega_p$, for each site can be used to enhance or remedy the effects of site inequivalence:
\begin{equation}\label{eq:balance}
\Omega_p \equiv z \frac{\Omega_k}{2} + \Omega_g.
\end{equation}
We can test if site pinning inequivalence is the mechanism preventing the $\alpha$--$(ET)_2 I_3$ lattice from having gaps. 
Indeed, reducing the precession frequencies of the sites with higher coordination numbers enables a band gap with chiral edge modes (\Fig{fig:alpha_lattice}B).
This provides another example of the inextricable connection between lattice geometry and topological order in gyroscopic lattices.

\begin{figure*}[t]
\includegraphics[width=\textwidth]{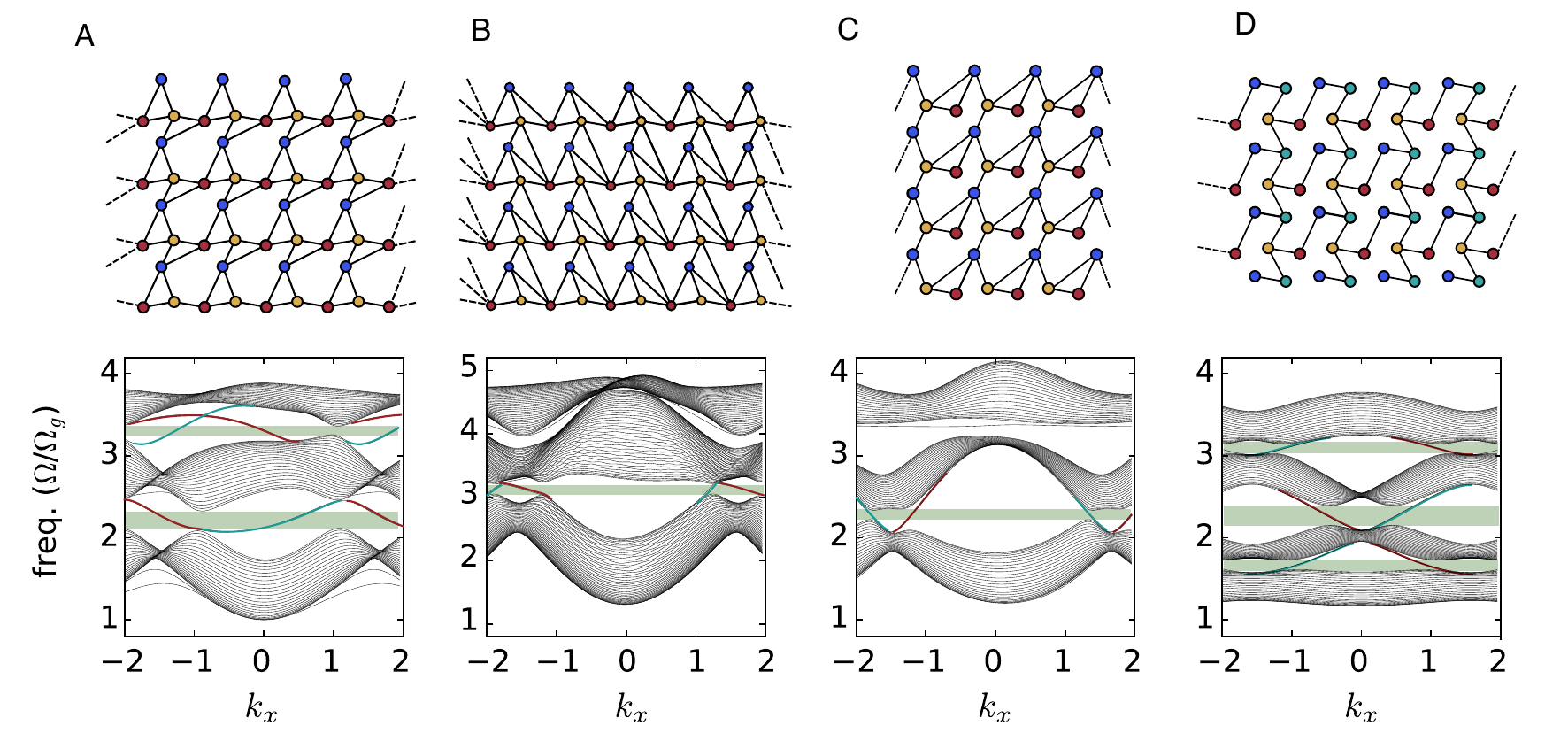}
\caption[]{\textbf{Examples of topological lattices created balancing coordination by varying on-site precession frequencies.}
Each lattice is generated by placing triangulated points in a square unit cell, then
deleting some bonds randomly.
\textit{(A)} An example of a deformed kagome lattice structure exhibits two topological gaps.
\textit{(B)} A mechanically stable lattice with one topological gap (upper gap) demonstrates that gyroscopic lattices need not be undercoordinated to be Chern insulators.
The propagation of edge modes is in the same direction as the kagome lattice. 
\textit{(C)} A 3-site-per-unit-cell lattice structure with one topological gap (lower gap).  The propagation of edge modes is in the same direction as the honeycomb lattice.
\textit{(D)} An example of a 4-site-per-unit-cell lattice structure with three topological gaps.  The propagation of edge modes is in the same direction as the kagome lattice for all three gaps.}
\label{lattice_examples}
\end{figure*}

\section{Towards topological design}
We have seen that both time-reversal symmetry and site equivalence are tied to lattice geometry and connectivity.
Turning now toward engineering new topological gyroscopic lattices, we can summarize the principles of the previous sections as follows:

\begin{enumerate}[I]
\item Breaking time reversal symmetry via bond angles is a necessary, but not sufficient condition for creating a lattice with a non-trivial band topology. 
\item A competition between time reversal symmetry and site equivalence determines whether or not a lattice can have topological modes.  
Lattice connectivity is relevant for determining the effective on-site precession frequencies to achieve equivalence.
\end{enumerate}
Using these two principles, one can construct topological metamaterials beginning with an arbitrary unit cell and subsequently balancing pinning frequencies according to Equation~\ref{eq:balance}.
This procedure can generate lattices with desired properties---such as multiple bandgaps or mechanical stability.
\Fig{lattice_examples} shows several examples. 

One example of a mechanically stable lattice with non-vanishing Chern number is shown in \Fig{lattice_examples}B. 
Although all previous lattices in this paper have been mechanically unstable ($\bar{z} \le 4$), the lattice in \Fig{lattice_examples}B shows that this is not necessary for band topology to arise.
Sublattices $A$ (yellow) and $C$ (red) have five bonds each, while sublattice $B$ (blue) has four. 
We expect that topological modes will arise when the total pinning at each site are approximately equal, which would occur for $\Omega_B > \Omega_{A,C}$. 
Figure 14 in the Supplementary Information shows that the numerics agree with this prediction~\cite{noauthor_notitle_nodate}.

The results demonstrated in this section show that topology is not specific to one family of lattices in gyroscopic networks and is in fact ubiquitous.
Many topological lattices can be created using only simple principles---opening a myriad possibilities for material design. 

\section{Conclusion}
In this article, we explored the interplay between lattice geometry and topological order in gyroscopic lattices --- including the effects of broken time reversal symmetry and site equivalence.
Along the way, we found examples of lattices with multiple band gaps containing edge modes of either chirality in the same structure and Chern numbers $|C| > 1$.
We then identified general principles which are helpful in designing lattices with desired topological band structures.
Building on our observations, we used a simple prescription that yields mechanically stable topological gyroscopic lattices and lattices with multiple band gaps.
The ubiquity of band topology in gyroscopic metamaterials provides a broad palette with which to design topological behaviors in elastic structures.
Further study could investigate the interplay between band topology and nonlinear excitations in gyroscopic networks or interspersing both clockwise and counterclockwise spinning sites.

\section{Acknowledgements}
We thank Ari Turner for providing the argument of Dirac cone protection in the canted bricklayer lattice, for detailed comments on the manuscript, and for useful discussions.
This work was primarily supported by the University of Chicago Materials Research Science and Engineering Center, which is funded by National Science Foundation under award number DMR-1420709. 
Additional support was provided by the  Packard Foundation. 
This work was also supported by NSF EFRI NewLAW grant 1741685.


\begin{thebibliography}{19}%
\makeatletter
\providecommand \@ifxundefined [1]{%
 \@ifx{#1\undefined}
}%
\providecommand \@ifnum [1]{%
 \ifnum #1\expandafter \@firstoftwo
 \else \expandafter \@secondoftwo
 \fi
}%
\providecommand \@ifx [1]{%
 \ifx #1\expandafter \@firstoftwo
 \else \expandafter \@secondoftwo
 \fi
}%
\providecommand \natexlab [1]{#1}%
\providecommand \enquote  [1]{``#1''}%
\providecommand \bibnamefont  [1]{#1}%
\providecommand \bibfnamefont [1]{#1}%
\providecommand \citenamefont [1]{#1}%
\providecommand \href@noop [0]{\@secondoftwo}%
\providecommand \href [0]{\begingroup \@sanitize@url \@href}%
\providecommand \@href[1]{\@@startlink{#1}\@@href}%
\providecommand \@@href[1]{\endgroup#1\@@endlink}%
\providecommand \@sanitize@url [0]{\catcode `\\12\catcode `\$12\catcode
  `\&12\catcode `\#12\catcode `\^12\catcode `\_12\catcode `\%12\relax}%
\providecommand \@@startlink[1]{}%
\providecommand \@@endlink[0]{}%
\providecommand \url  [0]{\begingroup\@sanitize@url \@url }%
\providecommand \@url [1]{\endgroup\@href {#1}{\urlprefix }}%
\providecommand \urlprefix  [0]{URL }%
\providecommand \Eprint [0]{\href }%
\providecommand \doibase [0]{http://dx.doi.org/}%
\providecommand \selectlanguage [0]{\@gobble}%
\providecommand \bibinfo  [0]{\@secondoftwo}%
\providecommand \bibfield  [0]{\@secondoftwo}%
\providecommand \translation [1]{[#1]}%
\providecommand \BibitemOpen [0]{}%
\providecommand \bibitemStop [0]{}%
\providecommand \bibitemNoStop [0]{.\EOS\space}%
\providecommand \EOS [0]{\spacefactor3000\relax}%
\providecommand \BibitemShut  [1]{\csname bibitem#1\endcsname}%
\let\auto@bib@innerbib\@empty
\bibitem [{\citenamefont {Thouless}\ \emph {et~al.}(1982)\citenamefont
  {Thouless}, \citenamefont {Kohmoto}, \citenamefont {Nightingale},\ and\
  \citenamefont {den Nijs}}]{thouless_quantized_1982}%
  \BibitemOpen
  \bibfield  {author} {\bibinfo {author} {\bibfnamefont {D.~J.}\ \bibnamefont
  {Thouless}}, \bibinfo {author} {\bibfnamefont {M.}~\bibnamefont {Kohmoto}},
  \bibinfo {author} {\bibfnamefont {M.~P.}\ \bibnamefont {Nightingale}}, \ and\
  \bibinfo {author} {\bibfnamefont {M.}~\bibnamefont {den Nijs}},\ }\href
  {\doibase 10.1103/PhysRevLett.49.405} {\bibfield  {journal} {\bibinfo
  {journal} {Phys. Rev. Lett.}\ }\textbf {\bibinfo {volume} {49}},\ \bibinfo
  {pages} {405} (\bibinfo {year} {1982})}\BibitemShut {NoStop}%
\bibitem [{\citenamefont {Prodan}\ and\ \citenamefont
  {Prodan}(2009)}]{prodan_topological_2009}%
  \BibitemOpen
  \bibfield  {author} {\bibinfo {author} {\bibfnamefont {E.}~\bibnamefont
  {Prodan}}\ and\ \bibinfo {author} {\bibfnamefont {C.}~\bibnamefont
  {Prodan}},\ }\href {\doibase 10.1103/PhysRevLett.103.248101} {\bibfield
  {journal} {\bibinfo  {journal} {Phys. Rev. Lett.}\ }\textbf {\bibinfo
  {volume} {103}},\ \bibinfo {pages} {248101} (\bibinfo {year}
  {2009})}\BibitemShut {NoStop}%
\bibitem [{\citenamefont {Rechtsman}\ \emph {et~al.}(2013)\citenamefont
  {Rechtsman}, \citenamefont {Zeuner}, \citenamefont {Plotnik}, \citenamefont
  {Lumer}, \citenamefont {Podolsky}, \citenamefont {Dreisow}, \citenamefont
  {Nolte}, \citenamefont {Segev},\ and\ \citenamefont
  {Szameit}}]{rechtsman_photonic_2013}%
  \BibitemOpen
  \bibfield  {author} {\bibinfo {author} {\bibfnamefont {M.~C.}\ \bibnamefont
  {Rechtsman}}, \bibinfo {author} {\bibfnamefont {J.~M.}\ \bibnamefont
  {Zeuner}}, \bibinfo {author} {\bibfnamefont {Y.}~\bibnamefont {Plotnik}},
  \bibinfo {author} {\bibfnamefont {Y.}~\bibnamefont {Lumer}}, \bibinfo
  {author} {\bibfnamefont {D.}~\bibnamefont {Podolsky}}, \bibinfo {author}
  {\bibfnamefont {F.}~\bibnamefont {Dreisow}}, \bibinfo {author} {\bibfnamefont
  {S.}~\bibnamefont {Nolte}}, \bibinfo {author} {\bibfnamefont
  {M.}~\bibnamefont {Segev}}, \ and\ \bibinfo {author} {\bibfnamefont
  {A.}~\bibnamefont {Szameit}},\ }\href
  {http://www.nature.com/nature/journal/v496/n7444/full/nature12066.html}
  {\bibfield  {journal} {\bibinfo  {journal} {Nature}\ }\textbf {\bibinfo
  {volume} {496}},\ \bibinfo {pages} {196} (\bibinfo {year}
  {2013})}\BibitemShut {NoStop}%
\bibitem [{\citenamefont {Nash}\ \emph {et~al.}(2015)\citenamefont {Nash},
  \citenamefont {Kleckner}, \citenamefont {Read}, \citenamefont {Vitelli},
  \citenamefont {Turner},\ and\ \citenamefont
  {Irvine}}]{nash_topological_2015}%
  \BibitemOpen
  \bibfield  {author} {\bibinfo {author} {\bibfnamefont {L.~M.}\ \bibnamefont
  {Nash}}, \bibinfo {author} {\bibfnamefont {D.}~\bibnamefont {Kleckner}},
  \bibinfo {author} {\bibfnamefont {A.}~\bibnamefont {Read}}, \bibinfo {author}
  {\bibfnamefont {V.}~\bibnamefont {Vitelli}}, \bibinfo {author} {\bibfnamefont
  {A.~M.}\ \bibnamefont {Turner}}, \ and\ \bibinfo {author} {\bibfnamefont
  {W.~T.~M.}\ \bibnamefont {Irvine}},\ }\href {\doibase
  10.1073/pnas.1507413112} {\bibfield  {journal} {\bibinfo  {journal}
  {Proceedings of the National Academy of Sciences}\ }\textbf {\bibinfo
  {volume} {112}},\ \bibinfo {pages} {14495} (\bibinfo {year}
  {2015})}\BibitemShut {NoStop}%
\bibitem [{\citenamefont {Mitchell}\ \emph
  {et~al.}(2018{\natexlab{a}})\citenamefont {Mitchell}, \citenamefont {Nash},\
  and\ \citenamefont {Irvine}}]{mitchell_realization_2018}%
  \BibitemOpen
  \bibfield  {author} {\bibinfo {author} {\bibfnamefont {N.~P.}\ \bibnamefont
  {Mitchell}}, \bibinfo {author} {\bibfnamefont {L.~M.}\ \bibnamefont {Nash}},
  \ and\ \bibinfo {author} {\bibfnamefont {W.~T.~M.}\ \bibnamefont {Irvine}},\
  }\href {\doibase 10.1103/PhysRevB.97.100302} {\bibfield  {journal} {\bibinfo
  {journal} {Phys. Rev. B}\ }\textbf {\bibinfo {volume} {97}},\ \bibinfo
  {pages} {100302} (\bibinfo {year} {2018}{\natexlab{a}})}\BibitemShut
  {NoStop}%
\bibitem [{\citenamefont {Mitchell}\ \emph
  {et~al.}(2018{\natexlab{b}})\citenamefont {Mitchell}, \citenamefont {Nash},
  \citenamefont {Hexner}, \citenamefont {Turner},\ and\ \citenamefont
  {Irvine}}]{mitchell_amorphous_2018}%
  \BibitemOpen
  \bibfield  {author} {\bibinfo {author} {\bibfnamefont {N.~P.}\ \bibnamefont
  {Mitchell}}, \bibinfo {author} {\bibfnamefont {L.~M.}\ \bibnamefont {Nash}},
  \bibinfo {author} {\bibfnamefont {D.}~\bibnamefont {Hexner}}, \bibinfo
  {author} {\bibfnamefont {A.~M.}\ \bibnamefont {Turner}}, \ and\ \bibinfo
  {author} {\bibfnamefont {W.~T.~M.}\ \bibnamefont {Irvine}},\ }\href {\doibase
  10.1038/s41567-017-0024-5} {\bibfield  {journal} {\bibinfo  {journal} {Nature
  Physics}\ }\textbf {\bibinfo {volume} {14}},\ \bibinfo {pages} {380}
  (\bibinfo {year} {2018}{\natexlab{b}})}\BibitemShut {NoStop}%
\bibitem [{\citenamefont {Wang}\ \emph {et~al.}(2015)\citenamefont {Wang},
  \citenamefont {Lu},\ and\ \citenamefont {Bertoldi}}]{wang_topological_2015}%
  \BibitemOpen
  \bibfield  {author} {\bibinfo {author} {\bibfnamefont {P.}~\bibnamefont
  {Wang}}, \bibinfo {author} {\bibfnamefont {L.}~\bibnamefont {Lu}}, \ and\
  \bibinfo {author} {\bibfnamefont {K.}~\bibnamefont {Bertoldi}},\ }\href
  {\doibase 10.1103/PhysRevLett.115.104302} {\bibfield  {journal} {\bibinfo
  {journal} {Physical Review Letters}\ }\textbf {\bibinfo {volume} {115}}
  (\bibinfo {year} {2015}),\ 10.1103/PhysRevLett.115.104302}\BibitemShut
  {NoStop}%
\bibitem [{\citenamefont {Kane}\ and\ \citenamefont
  {Lubensky}(2013)}]{kane_topological_2013}%
  \BibitemOpen
  \bibfield  {author} {\bibinfo {author} {\bibfnamefont {C.~L.}\ \bibnamefont
  {Kane}}\ and\ \bibinfo {author} {\bibfnamefont {T.~C.}\ \bibnamefont
  {Lubensky}},\ }\href
  {http://www.nature.com/nphys/journal/v10/n1/full/nphys2835.html} {\bibfield
  {journal} {\bibinfo  {journal} {Nature Physics}\ }\textbf {\bibinfo {volume}
  {10}},\ \bibinfo {pages} {39} (\bibinfo {year} {2013})}\BibitemShut {NoStop}%
\bibitem [{\citenamefont {S\"usstrunk}\ and\ \citenamefont
  {Huber}(2015)}]{susstrunk_observation_2015}%
  \BibitemOpen
  \bibfield  {author} {\bibinfo {author} {\bibfnamefont {R.}~\bibnamefont
  {S\"usstrunk}}\ and\ \bibinfo {author} {\bibfnamefont {S.~D.}\ \bibnamefont
  {Huber}},\ }\href {\doibase 10.1126/science.aab0239} {\bibfield  {journal}
  {\bibinfo  {journal} {Science}\ }\textbf {\bibinfo {volume} {349}},\ \bibinfo
  {pages} {47} (\bibinfo {year} {2015})}\BibitemShut {NoStop}%
\bibitem [{\citenamefont {Haldane}\ and\ \citenamefont
  {Raghu}(2008)}]{haldane_possible_2008}%
  \BibitemOpen
  \bibfield  {author} {\bibinfo {author} {\bibfnamefont {F.~D.~M.}\
  \bibnamefont {Haldane}}\ and\ \bibinfo {author} {\bibfnamefont
  {S.}~\bibnamefont {Raghu}},\ }\href
  {http://journals.aps.org/prl/abstract/10.1103/PhysRevLett.100.013904}
  {\bibfield  {journal} {\bibinfo  {journal} {Phys. Rev. Lett.}\ }\textbf
  {\bibinfo {volume} {100}},\ \bibinfo {pages} {013904} (\bibinfo {year}
  {2008})}\BibitemShut {NoStop}%
\bibitem [{\citenamefont {Ningyuan}\ \emph {et~al.}(2015)\citenamefont
  {Ningyuan}, \citenamefont {Owens}, \citenamefont {Sommer}, \citenamefont
  {Schuster},\ and\ \citenamefont {Simon}}]{ningyuan_time-_2015}%
  \BibitemOpen
  \bibfield  {author} {\bibinfo {author} {\bibfnamefont {J.}~\bibnamefont
  {Ningyuan}}, \bibinfo {author} {\bibfnamefont {C.}~\bibnamefont {Owens}},
  \bibinfo {author} {\bibfnamefont {A.}~\bibnamefont {Sommer}}, \bibinfo
  {author} {\bibfnamefont {D.}~\bibnamefont {Schuster}}, \ and\ \bibinfo
  {author} {\bibfnamefont {J.}~\bibnamefont {Simon}},\ }\href {\doibase
  10.1103/PhysRevX.5.021031} {\bibfield  {journal} {\bibinfo  {journal}
  {Physical Review X}\ }\textbf {\bibinfo {volume} {5}},\ \bibinfo {pages}
  {021031} (\bibinfo {year} {2015})}\BibitemShut {NoStop}%
\bibitem [{\citenamefont {Peano}\ \emph {et~al.}(2015)\citenamefont {Peano},
  \citenamefont {Brendel}, \citenamefont {Schmidt},\ and\ \citenamefont
  {Marquardt}}]{peano_topological_2015}%
  \BibitemOpen
  \bibfield  {author} {\bibinfo {author} {\bibfnamefont {V.}~\bibnamefont
  {Peano}}, \bibinfo {author} {\bibfnamefont {C.}~\bibnamefont {Brendel}},
  \bibinfo {author} {\bibfnamefont {M.}~\bibnamefont {Schmidt}}, \ and\
  \bibinfo {author} {\bibfnamefont {F.}~\bibnamefont {Marquardt}},\ }\href
  {\doibase 10.1103/PhysRevX.5.031011} {\bibfield  {journal} {\bibinfo
  {journal} {Physical Review X}\ }\textbf {\bibinfo {volume} {5}},\ \bibinfo
  {pages} {031011} (\bibinfo {year} {2015})}\BibitemShut {NoStop}%
\bibitem [{\citenamefont {Wang}\ \emph {et~al.}(2008)\citenamefont {Wang},
  \citenamefont {Chong}, \citenamefont {Joannopoulos},\ and\ \citenamefont
  {Soljačić}}]{wang_reflection-free_2008}%
  \BibitemOpen
  \bibfield  {author} {\bibinfo {author} {\bibfnamefont {Z.}~\bibnamefont
  {Wang}}, \bibinfo {author} {\bibfnamefont {Y.~D.}\ \bibnamefont {Chong}},
  \bibinfo {author} {\bibfnamefont {J.~D.}\ \bibnamefont {Joannopoulos}}, \
  and\ \bibinfo {author} {\bibfnamefont {M.}~\bibnamefont {Soljačić}},\
  }\href {\doibase 10.1103/PhysRevLett.100.013905} {\bibfield  {journal}
  {\bibinfo  {journal} {Phys. Rev. Lett.}\ }\textbf {\bibinfo {volume} {100}},\
  \bibinfo {pages} {013905} (\bibinfo {year} {2008})}\BibitemShut {NoStop}%
\bibitem [{\citenamefont {Fleury}\ \emph {et~al.}(2016)\citenamefont {Fleury},
  \citenamefont {Khanikaev},\ and\ \citenamefont {Alù}}]{fleury_floquet_2016}%
  \BibitemOpen
  \bibfield  {author} {\bibinfo {author} {\bibfnamefont {R.}~\bibnamefont
  {Fleury}}, \bibinfo {author} {\bibfnamefont {A.~B.}\ \bibnamefont
  {Khanikaev}}, \ and\ \bibinfo {author} {\bibfnamefont {A.}~\bibnamefont
  {Alù}},\ }\href {\doibase 10.1038/ncomms11744} {\bibfield  {journal}
  {\bibinfo  {journal} {Nature Communications}\ }\textbf {\bibinfo {volume}
  {7}},\ \bibinfo {pages} {11744} (\bibinfo {year} {2016})}\BibitemShut
  {NoStop}%
\bibitem [{\citenamefont {Haldane}(1988)}]{haldane_model_1988}%
  \BibitemOpen
  \bibfield  {author} {\bibinfo {author} {\bibfnamefont {F.~D.~M.}\
  \bibnamefont {Haldane}},\ }\href {\doibase 10.1103/PhysRevLett.61.2015}
  {\bibfield  {journal} {\bibinfo  {journal} {Phys. Rev. Lett.}\ }\textbf
  {\bibinfo {volume} {61}},\ \bibinfo {pages} {2015} (\bibinfo {year}
  {1988})}\BibitemShut {NoStop}%
\bibitem [{noa()}]{noauthor_notitle_nodate}%
  \BibitemOpen
  \href@noop {} {}\bibinfo {note} {See Supplemental Material at the end of this document. 
Supplementary Videos are available upon request.}\BibitemShut {Stop}%
\bibitem [{\citenamefont {Avron}\ \emph {et~al.}(1983)\citenamefont {Avron},
  \citenamefont {Seiler},\ and\ \citenamefont {Simon}}]{avron_homotopy_1983}%
  \BibitemOpen
  \bibfield  {author} {\bibinfo {author} {\bibfnamefont {J.~E.}\ \bibnamefont
  {Avron}}, \bibinfo {author} {\bibfnamefont {R.}~\bibnamefont {Seiler}}, \
  and\ \bibinfo {author} {\bibfnamefont {B.}~\bibnamefont {Simon}},\ }\href
  {\doibase 10.1103/PhysRevLett.51.51} {\bibfield  {journal} {\bibinfo
  {journal} {Phys. Rev. Lett.}\ }\textbf {\bibinfo {volume} {51}},\ \bibinfo
  {pages} {51} (\bibinfo {year} {1983})}\BibitemShut {NoStop}%
\bibitem [{\citenamefont {Laughlin}(1981)}]{laughlin_quantized_1981}%
  \BibitemOpen
  \bibfield  {author} {\bibinfo {author} {\bibfnamefont {R.~B.}\ \bibnamefont
  {Laughlin}},\ }\href {\doibase 10.1103/PhysRevB.23.5632} {\bibfield
  {journal} {\bibinfo  {journal} {Phys. Rev. B}\ }\textbf {\bibinfo {volume}
  {23}},\ \bibinfo {pages} {5632} (\bibinfo {year} {1981})}\BibitemShut
  {NoStop}%
\bibitem [{\citenamefont {Chaikin}\ and\ \citenamefont
  {Lubensky}(2000)}]{chaikin_principles_2000}%
  \BibitemOpen
  \bibfield  {author} {\bibinfo {author} {\bibfnamefont {P.~M.}\ \bibnamefont
  {Chaikin}}\ and\ \bibinfo {author} {\bibfnamefont {T.~C.}\ \bibnamefont
  {Lubensky}},\ }\href@noop {} {\emph {\bibinfo {title} {Principles of
  {Condensed} {Matter} {Physics}}}}\ (\bibinfo  {publisher} {Cambridge
  University Press},\ \bibinfo {year} {2000})\BibitemShut {NoStop}%
\bibitem [{\citenamefont {Ma\~nes}\ \emph {et~al.}(2007)\citenamefont
  {Ma\~nes}, \citenamefont {Guinea},\ and\ \citenamefont
  {Vozmediano}}]{Manes_existence_2007}%
  \BibitemOpen
  \bibfield  {author} {\bibinfo {author} {\bibfnamefont {J.~L.}\ \bibnamefont
  {Ma\~nes}}, \bibinfo {author} {\bibfnamefont {F.}~\bibnamefont {Guinea}}, \
  and\ \bibinfo {author} {\bibfnamefont {M.~A.~H.}\ \bibnamefont
  {Vozmediano}},\ }\href {\doibase 10.1103/PhysRevB.75.155424} {\bibfield
  {journal} {\bibinfo  {journal} {Phys. Rev. B}\ }\textbf {\bibinfo {volume}
  {75}},\ \bibinfo {pages} {155424} (\bibinfo {year} {2007})}\BibitemShut
  {NoStop}%
\bibitem [{\citenamefont {Wu}\ \emph {et~al.}(2015)\citenamefont {Wu},
  \citenamefont {Cho}, \citenamefont {Choi}, \citenamefont {Ge}, \citenamefont
  {Li}, \citenamefont {Han}, \citenamefont {Lubensky},\ and\ \citenamefont
  {Yang}}]{wu_directing_2015}%
  \BibitemOpen
  \bibfield  {author} {\bibinfo {author} {\bibfnamefont {G.}~\bibnamefont
  {Wu}}, \bibinfo {author} {\bibfnamefont {Y.}~\bibnamefont {Cho}}, \bibinfo
  {author} {\bibfnamefont {I.-S.}\ \bibnamefont {Choi}}, \bibinfo {author}
  {\bibfnamefont {D.}~\bibnamefont {Ge}}, \bibinfo {author} {\bibfnamefont
  {J.}~\bibnamefont {Li}}, \bibinfo {author} {\bibfnamefont {H.~N.}\
  \bibnamefont {Han}}, \bibinfo {author} {\bibfnamefont {T.}~\bibnamefont
  {Lubensky}}, \ and\ \bibinfo {author} {\bibfnamefont {S.}~\bibnamefont
  {Yang}},\ }\href {\doibase 10.1002/adma.201500716} {\bibfield  {journal}
  {\bibinfo  {journal} {Advanced Materials}\ }\textbf {\bibinfo {volume}
  {27}},\ \bibinfo {pages} {2747} (\bibinfo {year} {2015})}\BibitemShut
  {NoStop}%
\bibitem [{\citenamefont {Paulose}\ \emph {et~al.}(2015)\citenamefont
  {Paulose}, \citenamefont {Chen},\ and\ \citenamefont
  {Vitelli}}]{paulose_topological_2015}%
  \BibitemOpen
  \bibfield  {author} {\bibinfo {author} {\bibfnamefont {J.}~\bibnamefont
  {Paulose}}, \bibinfo {author} {\bibfnamefont {B.~G.-g.}\ \bibnamefont
  {Chen}}, \ and\ \bibinfo {author} {\bibfnamefont {V.}~\bibnamefont
  {Vitelli}},\ }\href {\doibase 10.1038/nphys3185} {\bibfield  {journal}
  {\bibinfo  {journal} {Nat Phys}\ }\textbf {\bibinfo {volume} {11}},\ \bibinfo
  {pages} {153} (\bibinfo {year} {2015})}\BibitemShut {NoStop}%
\bibitem [{\citenamefont {Sun}\ \emph {et~al.}(2012)\citenamefont {Sun},
  \citenamefont {Souslov}, \citenamefont {Mao},\ and\ \citenamefont
  {Lubensky}}]{sun_surface_2012}%
  \BibitemOpen
  \bibfield  {author} {\bibinfo {author} {\bibfnamefont {K.}~\bibnamefont
  {Sun}}, \bibinfo {author} {\bibfnamefont {A.}~\bibnamefont {Souslov}},
  \bibinfo {author} {\bibfnamefont {X.}~\bibnamefont {Mao}}, \ and\ \bibinfo
  {author} {\bibfnamefont {T.~C.}\ \bibnamefont {Lubensky}},\ }\href
  {http://www.pnas.org/content/109/31/12369} {\bibfield  {journal} {\bibinfo
  {journal} {Proc Natl Acad Sci USA}\ }\textbf {\bibinfo {volume} {109}},\
  \bibinfo {pages} {12369} (\bibinfo {year} {2012})}\BibitemShut {NoStop}%
\end{thebibliography}

%

\pagebreak 
\cleardoublepage
\appendix
\begin{titlepage}
   \vspace*{\stretch{1.0}}
   \begin{center}
      \Large\textbf{Supplementary Information for \\ `Tunable Band Topology in Gyroscopic Lattices'}\\
   \end{center}
   \vspace*{\stretch{2.0}}
\end{titlepage}


\section{Band structure and Chern number calculations}
Band topology in our gyroscopic lattices is encoded in the Chern number. 
To calculate the Chern number for gyroscopes on a lattice, we first find the band-structure using the linearized equation of motion, Eqn. 6 of the main text.  
For each site in a unit cell, we assume a solution which is composed of clockwise and counter-clockwise propagating modes: 
\begin{equation}
\label{sol1}
\psi_A = \psi_{A,R} e^{i(\vec{k} \cdot \vec{x} - \omega t)} + \bar{\psi}_{A,L} e^{-i(\vec{k} \cdot \vec{x} - \omega t)},
\end{equation}
where $A$ indexes the $n$ sites in a unit cell. 
The resulting equations can be expressed as the $2n \times 2n$ dynamical matrix which is a function of the wave vector, $\vec{k}$.
This dynamical matrix resembles hopping model matrices seen for lattice calculations in quantum mechanics. 

Diagonalizing the dynamical matrix yields $2n$ frequencies of the dispersion bands at each value of $\vec{k}$.
The eigenvalues come in positive/negative pairs, but each pair represents the same oscillation in real space. 
Because of this redundancy, we discuss only the positive eigenvalues for each system. 
At each value of $\vec{k}$, each mode has a corresponding eigenvector, $\big|u_j \big(\vec k\big)\big>$, characterizing the amplitudes and phases of the gyroscopes' collective motion.

The Chern number of each band is given by an integral of the Berry curvature, $\mathcal F(\vec k)$:
\begin{equation}
\label{chern}
\begin{split}
C_j &= \frac{1}{2\pi} \int \textrm{d}^2 k\ \mathcal F_j(\vec k) \\ 
& = \frac{i}{2 \pi} \oint A_j(k) \cdot \textrm{d} k, 
\end{split}
\end{equation}  
where $A_j(k) = i  \langle u_j | \nabla_k u_j\rangle $.
In this work, Chern numbers are calculated numerically using the phase-invariant formula~\cite{avron_homotopy_1983}
\begin{equation}
\label{chern_num}
C_{j} dx \wedge dy = \frac{i}{2 \pi} \int \textrm{d}^2 k\ \mathrm{Tr}[dP_j \wedge P_j dP_j],
\end{equation}
where $P_j \equiv |u_j \rangle \langle u_j |$ is the projection matrix and $\wedge$ is the wedge product.

\section{Time domain simulations}
We simulate our gyroscopic networks by numerically integrating the following equation with a fourth-order Runge-Kutta method:
\begin{equation}
\label{integration_eq}
\dot{\vec{\ell}} = \frac{\ell_f^2}{I \omega} (\vec{\ell_f} \times \vec{F}),
\end{equation}
where $\omega$ is the spinning frequency of the gyroscope, $I$ is the principal moment of inertia, $\ell_f$ denotes the distance from the pivot point to the position where the force, $\vec{F}$, acts. 
In our simulations, we consider only a gravitational restoring force and the effect of linear springs coupling neighbors in the lattice.

\section{Relationship to the Haldane model}
The lattice band structure maps our system to a tight-binding calculation in the limit where $\Omega_g \gg \Omega_k$.  
Although most calculations presented in this paper and previous experiments~\cite{nash_topological_2015,mitchell_amorphous_2018,mitchell_realization_2018} are far from this regime, it is useful for understanding the mechanism for topological band structures in gyroscopic lattices: if such a mapping does not close band gaps, then the topology of each band will not change.
Figure 4A of the main text shows that this assumption does not always hold: band gaps do close and reopen with changes in $\Omega_k/\Omega_g$ for the spindle lattice. 
However, for the honeycomb lattice, the gap width, $W$, is given by
\begin{equation}
	\label{eq:gapW}
	 \frac{W}{ \Omega_g} = \bigg(\frac{3}{2} R+1 - \sqrt{3R+ 1} \bigg),
\end{equation}
where $R= {\Omega_k}/{\Omega_g}$.  
In the limit of strong pinning ($\Omega_g \gg \Omega_k$), $W \approx (9/8) \Omega_k^2 / \Omega_g$.
Thus, for this particular case, the gap remains open, which explains the similarity of Figure 6A of the main text to the Haldane model phase diagram.

When $\Omega_g \gg \Omega_k$, the motion of each gyroscope is approximately circular and we may write the displacement of the gyroscope as $\psi_n = e^{-i \Omega t} u_n + e^{i \Omega t} v_n^*$, where $u_n$ is the amplitude of precession determined by gravity, $v_n$ is the counter-rotating amplitude, and $\Omega$ is the precession frequency of the normal mode.  
With the displacement written in this way, our equation of motion can be written as:
\begin{multline}\label{us}
\Omega u_n=\Omega_g u_n+\frac{1}{2}\Omega_k \sum_m (u_n-u_m)
\\
+\frac{1}{2}\Omega_k\sum_m (v_n-v_m)e^{2i\theta_{nm}},
\end{multline}
\begin{multline}\label{vs}
-\Omega v_n = \Omega_g v_n+\frac{1}{2}\Omega_k \sum_m (v_n-v_m)
\\ +\frac{1}{2}\Omega_k\sum_m (u_n-u_m)e^{-2i\theta_{nm}}.
\end{multline}

Assuming $\Omega_g \gg \Omega_k$, the gravitational frequency dominates, so that $|u_n| \gg |v_n|$ and all precession frequencies are close to the precession frequency: $\Omega \approx \Omega_g$.  
Because $|u_n| \gg |v_n|$, we may use equation \Eq{vs} to find an equation for $v_n$ in terms of $u_n$: $v_n\approx -\frac{\Omega_k}{4\Omega_g}\sum (u_n -u_m)e^{-2i\theta_{nm}}$, where the sum is over all neighbors $m$ of gyroscope $n$.  
We note that this term is the origin of the next-nearest neighbor coupling connecting our system to the Haldane model. 

When the result for $v_n$ in terms of $u_n$ is substituted into \Eq{us}, we obtain 
\begin{widetext}
\begin{equation}\label{eq:Haldane}
\Omega u_n=\Omega_g u_n+\frac{\Omega_k}{2} \sum_m (u_n-u_m)-\frac{\Omega_k^2}{8\Omega_g}\sum_{mm'} (u_n-u_m)e^{2i\theta_{m'nm}}+\frac{\Omega_k^2}{8\Omega_g}\sum_{ml}(u_m-u_l)e^{2i\theta_{nml}},
\end{equation}
\end{widetext}
where $\theta_{nml} = \theta_{nm}-\theta_{lm}$ is the angle between the bonds $nm$ and $lm$,
the second to last sum is over all pairs of neighbors $m$ and $m'$ of $n$, and the last sum is over all
neighbors $m$ of $n$ and neighbors $l$ of $m$. 
Here, our last term corresponds to next-nearest neighbor coupling.

To compare directly with the topological Haldane model, we will analyze the lowest-order terms for the sums over the nearest neighbors and next-nearest neighbors: 
\begin{multline}\label{eq:Haldane_mt}
\Omega u_n = \Omega_g u_n+\sum_m (u_n-u_m) + \\
\frac{\Omega_k^2}{8\Omega_g}\sum_{m,l}(u_m-u_l)e^{2i\theta_{nml}}.
\end{multline}
$\Omega_k$ is similar to Haldane's nearest neighbor tunneling amplitude, and ${\Omega_k^2}/{8 \Omega_g}$ is the next-nearest neighbor tunneling amplitude.
As in Haldane's model, the next-nearest neighbor coupling terms have a complex exponential. 
In tight binding systems, this phase factor is calculated using the Peierls substitution as  $\theta = \int \vec{A}
\cdot d\vec{\ell}$, where $d\vec{\ell}$ is along the tunneling path. 
In gyroscopic systems, the phase factor is governed purely by the lattice geometry.

\section{Topological phase transitions in gyroscopic lattices}

\begin{figure}
\includegraphics[width=\columnwidth]{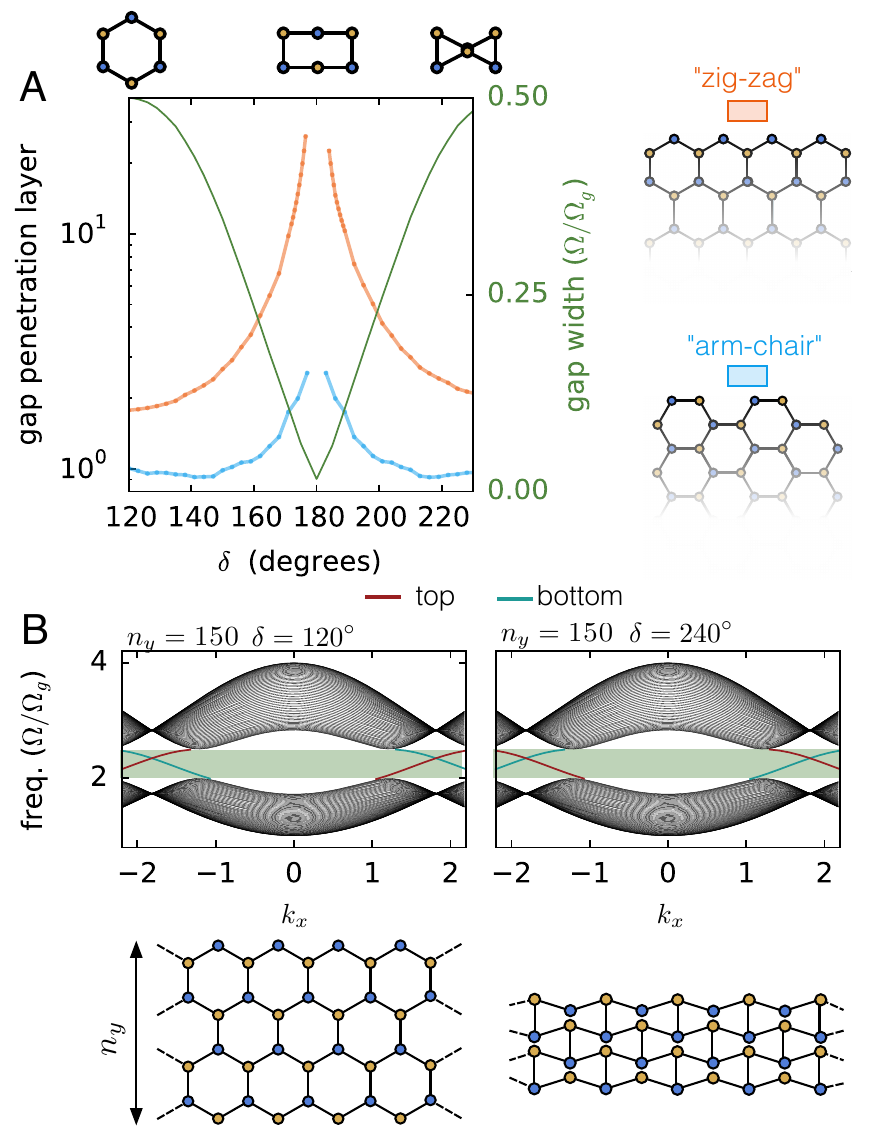}
\caption[]{\textbf{Closing the gap, diverging length scales, and flipped edge modes} 
\textit{(A)} The minimum penetration depth of the edge modes increases with decreasing gap with for both zig-zag (orange) and arm-chair (blue) edges in the honeycomb lattice.
\textit{(B)} The right-going excitations are on the top of the strip when $\delta < 180 ^{\circ}$ and bottom for $\delta > 180^{\circ}$. 
}
\label{penetration_depth}
\end{figure}

Previous work studied the topological phase transition that occured when the honeycomb lattice was deformed through a bricklayer transition into a `bowtie' lattice~\cite{nash_topological_2015}.
In this work, we study this transition with an additional parameter, $\phi$, and note its similarities with phase transitions in electronic systems. 

As shown in \Fig{penetration_depth}A, as the lattice is deformed towards the bricklayer lattice, the gap width decreases and minimum penetration depth of the edge modes diverges for both a zig-zag and arm-chair boundary edge.  
The penetration depths shown in this plot were found using normal mode calculations in long strips with both zig-zag and arm-chair boundaries. The penetration depths were taken as the value $\xi$ when the mode amplitudes were fit with a decaying exponential $|\psi| \sim e^{-x/\xi}$, where $x$ is the minimum number of nearest neighbor hoppings of a site from the edge.

\Fig{penetration_depth}B indicates that the propagation direction indeed changes when the Chern number changes.  
In the left panel of \Fig{penetration_depth}B, the modes on the top of the network have a positive group velocity, indicating a clockwise edge excitation propagation direction.  On the right, in the bowtie configuration, the direction has reversed.

\section{Dirac cone protection in the canted bricklayer lattice}

Surprisingly, Dirac points are preserved in the canted bricklayer configuration, despite the fact that shearing the bricklayer configuration breaks time reversal symmetry.
While time reversal symmetry is broken, a combination of time reversal and discrete symmetries can stabilize the Dirac points~\cite{Manes_existence_2007}. 
Here, a peculiar kind of reflection in the canted bricklayer lattice creates a line in momentum space along which decoupled states cross, leading to a Dirac point which may move in $k$ but cannot become gapped.

\begin{figure}
\includegraphics[width=90mm]{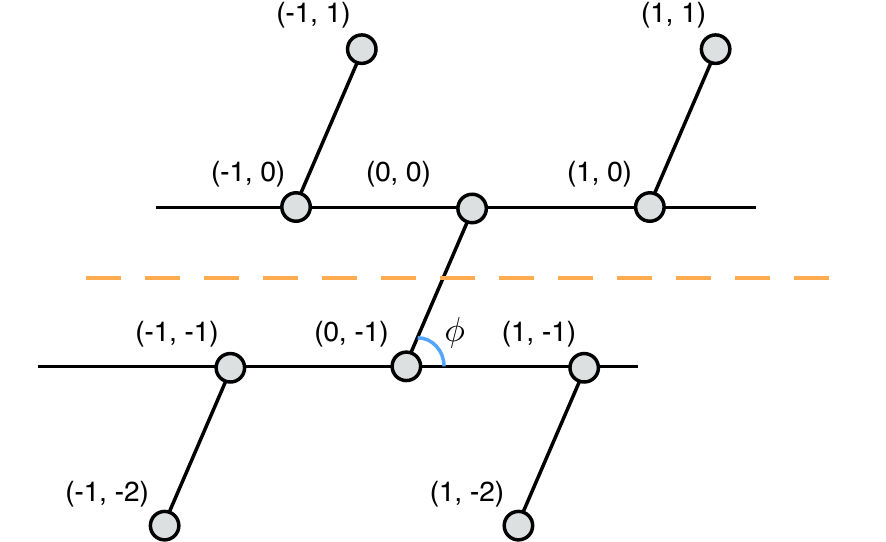}
\caption{
\textbf{A pseudo-reflection symmetry in the canted bricklayer takes site $(n,m)$ to site $(n, -1-m)$.} 
The symmetry is a reflection across the orange dashed line, combined with a shearing of space to preserve the tilt angle, $\phi$.
}
\label{fig_schematic}
\end{figure}

\subsection{Symmetry of the equations of motion}
To see the pseudo-reflection symmetry, we index gyroscopes by the label $(n,m)$ as shown in \Fig{fig_schematic} 
It is useful to work in the $(x,y)$ basis, so that the equations of motion are
\begin{align}
\begin{split}
\dot{x}_{n,m} = & \Omega_gy_{n,m} + \Omega_k 
(x_{n,m} - x_{n, m\pm 1})\cos \phi \sin \phi \\
& + \Omega_k (y_{n,m} - y_{n,m\pm 1}) \sin^2 \phi ,
\end{split}
\\
\begin{split}
\dot{y}_{n,m} = & -\Omega_g x_{n,m} 
\\ & +\Omega_k \left( x_{n+1,m} + x_{n-1, m} - 2 x_{n,m}
\right)
\\ & + \Omega_k 
 (x_{n,m\pm 1} - x_{n,m}) \cos^2 \phi 
 \\
 &  +  \Omega_k (y_{n,m\pm 1} - x_{n, m}) \cos \phi \sin \phi ,
\end{split}
\end{align}
where the plus and minus signs correspond to $(n+m)$ being odd or even, respectively.
The equations of motion exhibit a symmetry under the exchange $(n,m) \rightarrow (n, -1-m)$ --- that is, if $\pvec{r}'_{n,m} \equiv \vec{r}_{n, -1-m} = (x_{n, -1-m}, y_{n, -1-m})$, then $\pvec{r}'$ satisfies the same equations as $\vec{r}$.
To prove this, we apply the original equations with $(n, -1-m)$ and note that if $m$ was even, then $-1-m$ is odd, so that the $\pm$ signs are taken to $\mp$, and the result follows.
This symmetry exchanges any pair of connected gyroscopes with another pair that are connected by a spring with the same slope, so we immediately find that the energy is invariant under this symmetry.
This is only possible when the action of the symmetry preserves both the connectivity of the lattice and preserves the directions of bonds up to rotations by 180 degrees. 
As a result, among the family of deformed honeycomb lattice, this pseudo-reflection symmetry arises only for the bricklayer and canted bricklayer geometries.

\subsection{Symmetry of Bloch waves}
The pseudo-reflection symmetry acts in real space, but also transforms momentum space.
By Bloch's theorem, $\vec{r}_{n,m} = e^{i \vec{k} \cdot \vec{R}_{n,m}} u$, where $u$ takes one value for $A$ sublattice sites and another for $B$ sublattice sites. 
It follows that 
\begin{equation}
\pvec{r}'_{n,m} = \vec{r}_{n,-1-m} \propto e^{i \vec{k} \cdot \vec{R}_{n, -1-m}}.
\end{equation}
If we define $\vec{R}_{n,m} = (n  + m \cos \phi, m \sin \phi)  \equiv (x_{n,m}, y_{n,m})$,
then applying $(n,m) \rightarrow (n, -1-m)$ gives
\begin{align}
\pvec{R}'_{n,-1-m} =& (n - (m + 1) \cos \phi, -(m + 1)  \sin \phi)  \nonumber
\\ 
=& (x_{n,m} - 2 y_{n,m} \cot \phi - \cos \phi, 
\\& -y_{n,m} -\sin \phi). \nonumber
\end{align} 
Setting $\pvec{r}'_{n,m} = \exp{[i k'\cdot R_{n,m}]}$ up to a phase factor which is independent of $n$ and $m$ provides a correspondence between $\pvec{k}'$ and $\vec{k}$:
\eq
\pvec{k}' = (k_x, -2 \cot \phi - k_y).
\qe

Choosing our unit cell to contain $(n,m) = (0, 0)$ and $(0, -1)$, we denote the degrees of freedom
\eq
\left( \begin{array}{cc}
x_{0,0} & y_{0,0} \\
x_{0,-1} & y_{0, -1}
\end{array}
\right) = 
\left(  \begin{array}{cc}
A & B \\
C & D 
\end{array}
\right).
\qe
The pseudo-reflection symmetry operation, now defined as 
\eq 
A,B \leftrightarrow C,D \textrm{ and } k_y \rightarrow -2 \cot \phi k_x - k_y 
\qe
 takes $\dot{A} \leftrightarrow \dot{C}$ and $\dot{B} \leftrightarrow \dot{D}$.
 
 \subsection{Stability of Dirac points}
We now show that the pseudo-reflection symmetry ($A,B \leftrightarrow C,D$ and $k_y \rightarrow -2 \cot \phi k_x - k_y $) fixes a line in momentum space on which modes can be classified as even and odd under the action of the reflection. 
In general, a band crossing can be avoiding due to hybridization (i.e. a coupling between the bands). 
However, the even and odd bands which arise are decoupled and can therefore provide a protected crossing. 

\begin{figure}
\includegraphics[width=90mm]{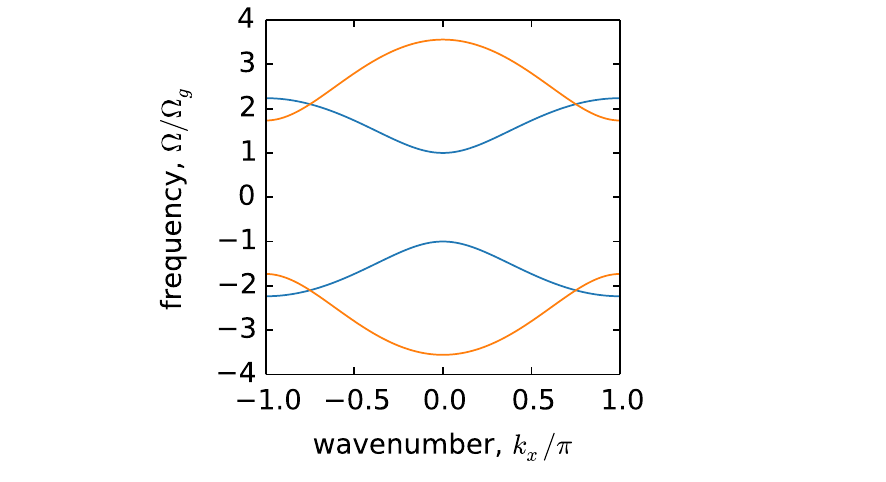}
\caption{
\textbf{The dispersion of the canted bricklayer geometry has protected Dirac points.} 
Along the line $k_y = -\cot \phi k_x$ in momentum space, symmetric and antisymmetric modes decouple. 
The modes that are symmetric (antisymmetric) under pseudo-reflection symmetry are shown as blue (orange) curves. 
Since the symmetric and anti-symmetric modes cross, Dirac cones appear in the spectrum, even though time reversal symmetry is broken.
Here, we set $\Omega_g = \Omega_k$ and the tilt angle $\phi = 1 $ radian. }
\label{fig_ABCD}
\end{figure}

Consider the wavevectors $\vec{k}$ such that $\vec{k} = \vec{k}' + \vec{G}$, where $\vec{G}$ is either zero or a reciprocal lattice vector.
This condition defines a line of momenta on which the equations are already symmetric under $(A, B) \rightarrow (C,D)$.
Now, the variables
$(A + C, B+D)$ and $(A-C, B-D)$ are symmetric and antisymmetric under the pseudo-reflection symmetry, which implies that we expect them to be decoupled from each other.
Below we compute the dispersions for each set of variables separately along the line defined by $\vec{k}' - \vec{k} = 0$. 
These equations are indeed uncoupled from each other, so any crossings that arise cannot open into gaps.

Choosing Bravais lattice vectors $\vec{R} = (R_x, R_y) = (2,0)$ and $(1 + \cos \phi, \sin \phi)$, we have reciprocal lattice vectors
\eq 
\vec{G} = n \left(0, \frac{2\pi}{\sin \phi} \right) +
 m \pi  \left(1,
\frac{-1-\cos\phi}{\sin \phi}  \right).
\qe 
We can seek curves in momentum space such that $k$ and $\pvec{k}'$ differ by only a reciprocal lattice vector, $\pvec{k}' - \vec{k} = \vec{G}$. 
Enforcing $\pvec{k}' - \vec{k} = (0, -2 \cot \phi k_x - 2 k_y) = \vec{G}$ constrains $m=0$, and $k_y = \frac{-n \pi }{\sin \phi} - \cot \phi k_x$.
These define a set of parallel lines separated by a reciprocal lattice vector.
Since all curves are equivalent, we may use $n=0$ without loss of generality.
The result is
\begin{align} \label{eqmotkAC}
\dot{A} + \dot{C} = & \Omega_g (B + D) 
\\
\begin{split} \label{eqmotkA-C}
\dot{A} - \dot{C}  = & \Omega_g (B-D) + 
2 \Omega_k (B-D) \sin^2\phi 
\\ &+ 2 \Omega_k (A-C) \cos \phi \sin \phi  ]
\end{split}
\\
\label{eqmotkBD}
\dot{B} + \dot{D} = & (2\Omega_k \cos k_x -\Omega_g  - 2\Omega_k)(A + C)  
\\
\begin{split}
\label{eqmotkB-D}
\dot{B} - \dot{D} = & (-2\Omega_k \cos k_x -\Omega_g A - 
2 \Omega_k)(A-C) \\
& - 2 \Omega_k [\cos^2 \phi (A-C) + \cos \phi \sin \phi (B-D)].
\end{split}
\end{align}
The eigenvalues are plotted in \Fig{fig_ABCD}, and the corresponding eigenvectors decouple into even modes (depending only on $A+C$ and $B+D$) and odd modes (depending only on $A-C$ and $B-D$).
This protects the band crossings which form the Dirac points.

We note that the same conclusion would follow in the case where bond lengths are unequal. 
Even more generally, this symmetry arises for any configuration in which the action of the pseudo-reflection both preserves the connectivity of the lattice and preserves the directions of bonds up to 180 degrees.

\section{Procedure for generating topological lattices}
The lattices shown in Figure 8 of the main text were generated through repeated application of the procedure outlined here. 
Lattices were created by placing points in a square unit cell, repeating, and triangulating.  
A random number of bonds in the unit cell were then deleted.
The gravitational  precession frequencies were then balanced in accordance with Eqn. 8 in the main text using the coordination number $z$  of each site.

\begin{figure}
\includegraphics[width=\columnwidth]{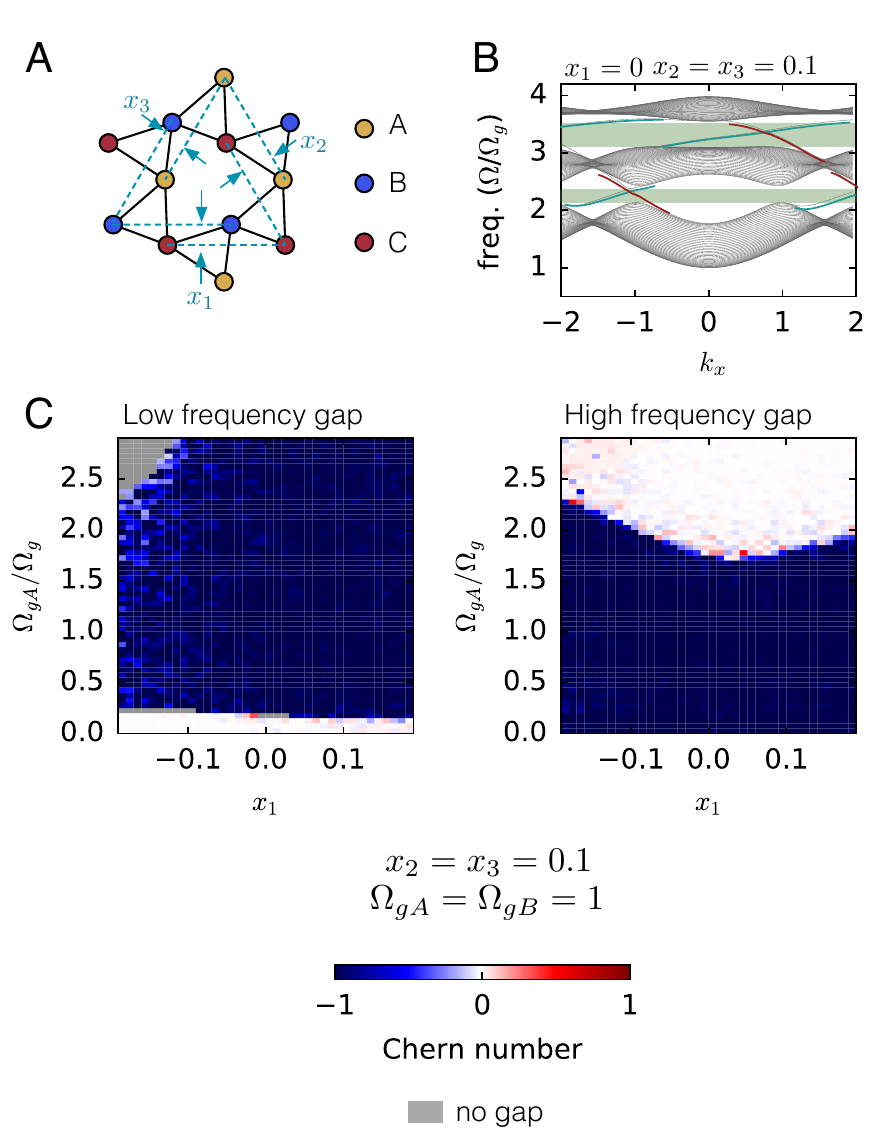}
\caption[]{\textbf{Kagome lattice phase diagram and band structure}.  \textit{(A)} We consider deformations of the kagome lattice using geometry parameters $x_1, x_2, x_3$. 
\textit{(B)} A calculation on a strip which is infinite along $x$ (as depicted in \textit{A}) shows that both gaps have edge modes which propagate in the same direction on the top edges.  
\textit{(C)} The sum of band Chern numbers for bands below each gap is shown in the phase diagrams for each of the two gaps. 
The Chern number has only a weak dependence on lattice geometry near the transition in the high frequency band gap. 
}
\label{kagome_first}
\end{figure}

\begin{figure}
\includegraphics[width=\columnwidth]{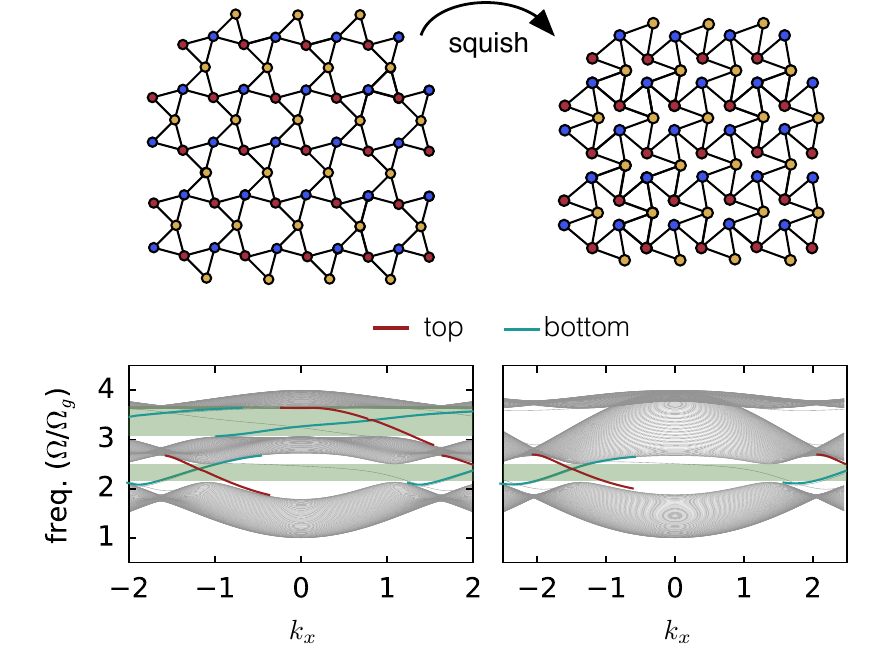}
\caption[]{
\textbf{Controlling the gap width in a kagome lattice via global deformations.} 
\textit{(A)} A twisted kagome lattice can be squished on the edges to create a uniformly collapsing material. 
\textit{(B)} The upper band gap closes when this operation is performed.  
This shows an example of an easily tunable band gap material.  }

\label{kagome_collapse}
\end{figure}

\section{The deformed kagome lattice} 
Here we explore the topology of the deformed kagome lattice, which can be constructed such that the only non-localized zero energy mode is a uniformly collapsing mode~\cite{wu_directing_2015}.
The lattice --- which has three sites per unit cell --- also has the property of having two topologically nontrivial band-gaps.

We note this lattice has been found to have interesting topological properties in other mechanical systems with time reversal symmetry.
In mechanical frames, topological modes have been experimentally proposed and demonstrated for kagome and deformed kagome lattices with variation of lattice geometry~\cite{kane_topological_2013,paulose_topological_2015}. 
In these cases,  variation of lattice geometry leads to a topological polarization which preferentially concentrates zero modes or states of self stress on one side of the material or at dislocations.

In order to investigate the topology of the kagome lattice under geometric deformations for the gyroscopic case, we employ the lattice distortion shown in \Fig{kagome_first}A. 
As in~\cite{kane_topological_2013}, the primitive vectors of the lattice are fixed as $a_{p+1} = (\mathrm{cos}(2 \pi p/3), \mathrm{sin}(2\pi p/3))$.
The positions of the lattice sites are described as $d_1 = a_1/2 + s_2$, $d_2 = a_2/2 - s_2$ and  $d_3 = a_3/2$, where $s_p$ describes the displacement of $d_{p-1}$ relative to the midpoint of the line along $a_p$.
Following~\cite{kane_topological_2013}, these deformations can be represented by the independent variables $x_1, x_2, x_3$, and $z$ 
so that $s_p = x_p (a_{p-1} - a_{p+1}) + \gamma_{p} a_{p}$ with $z = \gamma_1 + \gamma_2 + \gamma_3$.

The three bands and two band gaps in the vibrational spectrum of this lattice are shown in green in \Fig{kagome_first}B for equivalent sites ($\Omega_{gA} = \Omega_{gB} = \Omega_{gC}$) and geometry parameters $x_1=0, x_2=x_3=0.1$.
The Chern number of the lowest band is opposite that of the lowest band in the honeycomb lattice for the same gyroscope spin direction.  
The middle band has a Chern number of 0, and the top band has a Chern number opposite the bottom band.

Variation of lattice geometry has only a slight effect on the topology of the band structure, as shown in \Fig{kagome_first}C.  
In fact, for a lattice with equivalent lattice sites, the geometry variation has no effect on the system's topology.
These results persist for variations of any single precession frequency and geometry parameter.

While lattice topology exhibits only a slight dependence on changes in a single lattice geometry parameter, the upper band gap can additionally be closed via simultaneous adjustment of all three lattice parameters, corresponding to twisting the kagome lattice~\cite{sun_surface_2012}.
When the kagome lattice is in a pre-twisted state ($x_1 = x_2 = x_3 \neq 0$), a uniformly collapsing mode is admitted if a force is applied at the boundary (top panel of \Fig{kagome_collapse})~\cite{wu_directing_2015}. 
Using the parameterization above, the collapsed twisted kagome can be achieved by increasing the absolute value of $x_1, x_2, x_3$ and adjusting the overall scale to prevent changes in the rest bond length.

The width of the upper band gap may be controlled by adjustment of the degree of collapse of the lattice.
As shown in \Fig{kagome_collapse}, a slightly twisted lattice 
has the same three bands and two band-gaps as in \Fig{kagome_first}B.
However, the second band gap is closed as the degree of twist increases
(\Fig{kagome_collapse} right).
We therefore find that one set of edge states in a twisted kagome metamaterial can be conveniently controlled by applying or releasing stress on the boundary.

\section{A mechanically stable lattice with nontrivial band topology}

\begin{figure}
\includegraphics[width=\columnwidth]{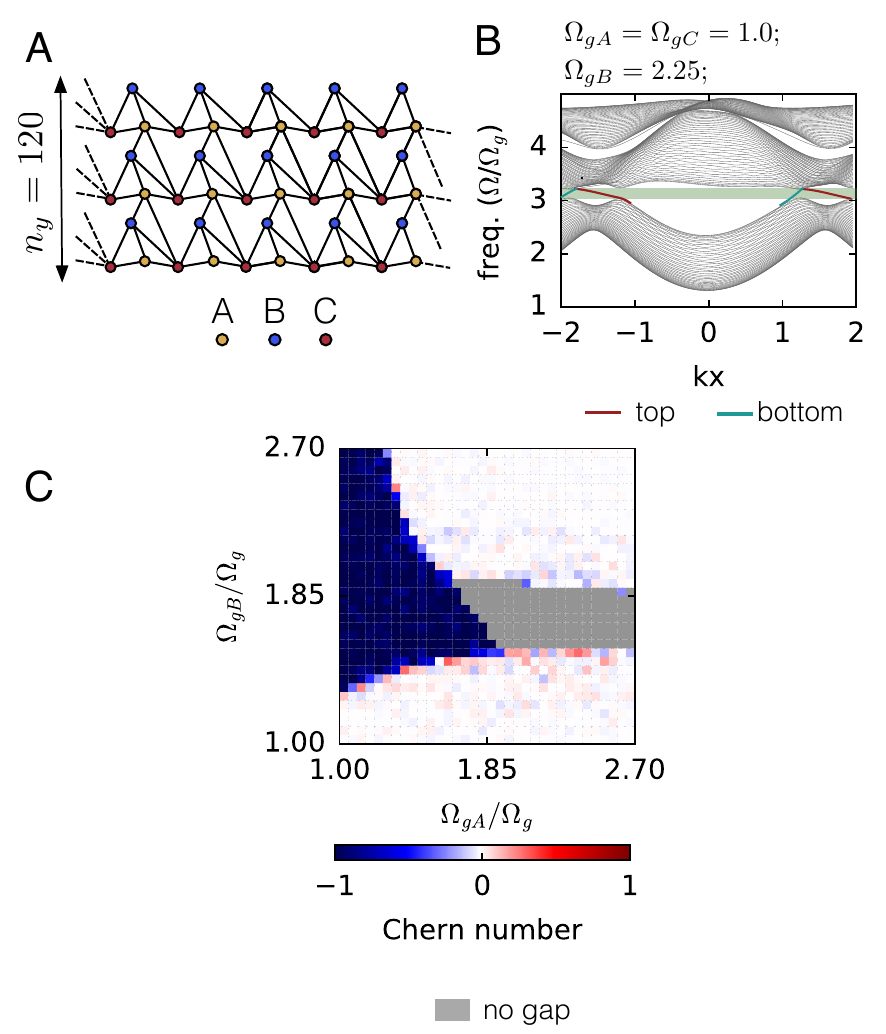}
\caption[]{
\textbf{Mechanically stable lattice with topologically nontrivial band structure.}   
\textit{(A)} A semi-infinite strip of an example of a mechanically stable lattice with 3 sites in each unit cell. $z_A = 5$, $z_B = 4$, $z_C = 5$.
\textit{(B)} The band structure of the semi-infinite strip shown in \textit{(A)} 
\textit{(C)} The structure is topologically nontrivial when the pinning frequency for the B sites (the sites with the lowest coordination number) is higher than $\Omega_{gA}$ and $\Omega_{gC}$, as expected from Equation 8 of the main text.
}
\label{oc_lattice}
\end{figure}
Over-coordinated ($\bar{z} > 4$), mechanically stable lattices can also support chiral edge modes.  
One such example, shown in the main text, is described in more detail in \Fig{oc_lattice}.
When the on-site pinning (Eqn 8. in main text) is balanced for the lattice, one of the band-gaps supports topological excitations \Fig{oc_lattice}B. 
The phase diagram, \Fig{oc_lattice}C, confirms that there is a neighborhood around this value for which topological excitations occur. 


\end{document}